\documentclass[aps,prd,twocolumn,superscriptaddress,amsmath,amssymb,nofootinbib,longbibliography,preprintnumbers,floatfix]{revtex4-2}
\usepackage[dvipsnames]{xcolor}
\usepackage[colorlinks]{hyperref}
\usepackage{xspace}
\usepackage{mathtools}
\usepackage{bm}
\usepackage{makecell}

\usepackage{tikz}
\usetikzlibrary{tikzmark}
\usetikzlibrary{positioning}
\usepackage[normalem]{ulem} 

\hypersetup{linkcolor=BrickRed,citecolor=Green,
filecolor=Mulberry,
urlcolor=NavyBlue,
menucolor=BrickRed,
runcolor=Mulberry
}

\definecolor{lime}{HTML}{A6CE39}
\DeclareRobustCommand{\orcidicon}{\hspace{-1mm}
	\begin{tikzpicture}
	\draw[lime, fill=lime] (0,0) 
	circle [radius=0.16] 
	node[white] {{\fontfamily{qag}\selectfont \tiny \,ID}};
	\draw[white, fill=white] (-0.0525,0.095) 
	circle [radius=0.007];
	\end{tikzpicture}
	\hspace{-3mm}
}

\foreach \x in {A, ..., Z}{\expandafter\xdef\csname orcid\x\endcsname{\noexpand\href{https://orcid.org/\csname orcidauthor\x\endcsname}
			{\noexpand\orcidicon}}
}

\usepackage{graphicx}
\usepackage{bbm}

\makeatletter
\newcommand{\abs}{\@ifstar\abssmall\absbig}
\newcommand{\absbig}[1]{\left \lvert #1 \right \rvert}
\newcommand{\abssmall}[1]{\lvert #1 \rvert}
\makeatother

\newcommand{\x}{\mathrm{x}}
\newcommand{\y}{\mathrm{y}}
\newcommand{\z}{\mathrm{z}}
\newcommand{\ii}{\mathrm{i}}

\renewcommand{\Im}{\mathrm{Im}}
\renewcommand{\Re}{\mathrm{Re}}
\newcommand{\pdv}[2]{\frac{\partial #1}{\partial #2}}

\newcommand{\dd}{\mathrm{d}}

\newcommand{\bN}{\overline{N}}
\newcommand{\bF}{\overline{F}}
\newcommand{\bP}{\overline{P}}
\newcommand{\bA}{\bar{A}}
\newcommand{\bB}{\overline{B}}
\newcommand{\bC}{\overline{C}}

\newcommand{\bDelta}{\overline{\Delta}}

\newcommand{\bGamma}{\overline{\Gamma}}

\newcommand{\vrho}{\varrho}
\newcommand{\bvrho}{\bar{\varrho}}
\newcommand{\N}{\mathcal{N}}
\newcommand{\F}{\mathcal{F}}

\newcommand{\appsection}[1]{\section{\MakeUppercase{#1}}}

\renewcommand{\vec}{\mathbf}

\begin{document}
\preprint{N3AS-25-020}

\title{Neutrino flavor instabilities in neutron star mergers with moment transport: Slow, fast, and collisional modes}

\author{Julien Froustey\orcidA{}}
\email{julien.froustey@ific.uv.es}
\affiliation{Institut de Física Corpuscular (CSIC-Universitat de València), Parc Científic UV, C/ Catedrático José Beltrán 2, E-46980 Paterna (Valencia), Spain}

\author{Francois Foucart\orcidF{}}
\affiliation{Department of Physics \& Astronomy, University of New Hampshire, 9 Library Way, Durham, NH 03824, USA}

\author{Christian Hall\orcidC{}}
\affiliation{Department of Physics and Astronomy, North Carolina State University, Raleigh, NC 27695, USA}

\author{James P. Kneller\orcidJ{}}
\affiliation{Department of Physics and Astronomy, North Carolina State University, Raleigh, NC 27695, USA}

\author{Debraj Kundu\orcidD{}}
\affiliation{Department of Physics and Astronomy, University of Tennessee Knoxville, Knoxville, TN 37996, USA}

\author{Zidu Lin\orcidZ{}}
\affiliation{Department of Physics \& Astronomy, University of New Hampshire, 9 Library Way, Durham, NH 03824, USA}
\affiliation{Department of Physics and Astronomy, University of Tennessee Knoxville, Knoxville, TN 37996, USA}
\affiliation{Sino-French Institute of Nuclear Engineering and Technology, Sun Yat-Sen University, Zhuhai 519082, China}

\author{Gail C. McLaughlin\orcidG{}}
\affiliation{Department of Physics and Astronomy, North Carolina State University, Raleigh, NC 27695, USA}

\author{Sherwood Richers\orcidS{}}
\affiliation{Department of Physics and Astronomy, University of Tennessee Knoxville, Knoxville, TN 37996, USA}

\begin{abstract}
    Determining where, when, and how neutrino flavor oscillations must be included in large-scale simulations of hot and dense astrophysical environments is an enduring challenge that must be tackled to obtain accurate predictions. Using an angular moment-based linear stability analysis framework, we examine the different kinds of flavor instabilities that can take place in the context of the post-processing of a neutron star merger simulation, with a particular focus on the collisional flavor instability and a careful assessment of several commonly used approximations. First, neglecting anisotropies of the neutrino field, we investigate the extent to which commonly used monoenergetic growth rates reproduce the results obtained from a full multi-energy treatment. Contrary to the large discrepancies found in core-collapse supernova environments, we propose a simple combination of energy-averaged estimates that reproduces the multi-energy growth rates in our representative simulation snapshot. We then quantify the impact of additional physical effects, including nuclear many-body corrections, scattering opacities, and the inclusion of the vacuum term in the neutrino Hamiltonian. Finally, we include the neutrino distribution anisotropies, which allows us to explore, for the first time in a multi-energy setting, the interplay between collisional, fast, and slow modes in a moment-based neutron star merger simulation. We find that, despite a dominance of the fast instability in most of the simulation volume, certain regions exhibit only a collisional instability, while others, especially at large distances, exhibit a slow instability that is largely underestimated if anisotropic effects are neglected.
\end{abstract}

\maketitle

%%%%%%%%%%%%%%%%%%%%%%
\section{Introduction}
%%%%%%%%%%%%%%%%%%%%%%

In the era of multimessenger astronomy, the modeling of hot and dense astrophysical environments such as core-collapse supernovae (CCSNe) and neutron star mergers (NSMs) relies on increasingly sophisticated large-scale simulations that incorporate a broad range of physical processes, including (magneto)hydrodynamics, general relativity, and detailed nuclear microphysics~\cite{Janka:2016fox,Muller:2020ard,Burrows:2020qrp,Mezzacappa:2020pkk,Kyutoku:2021icp,Radice:2020ddv,Radice:2024gic}. Neutrino transport plays a key role in these systems, influencing their dynamics, energetics, and nucleosynthesis~\cite{Mezzacappa:2020oyq,Foucart:2022bth,Fischer:2023ebq,Wang:2023tso}. Despite significant progress in the realism of state-of-the-art simulations, neutrino flavor oscillations are not consistently included. However, it is now widely recognized that these environments harbor neutrino \emph{flavor instabilities}, which lead to to an exponential growth of flavor coherence (a measure of the superposition of flavor states) starting from quasi-pure flavor states (see~\cite{Capozzi_review,Tamborra_review,Volpe:2023met,Johns:2025mlm} for recent reviews). Different mechanisms can produce flavor instabilities. Historically, so-called ``slow modes," driven by the neutrino mass differences, were investigated in astrophysical and early Universe environments (see~\cite{Duan_review,2009JPhG...36k3201D} and references therein). After the initial observation that anisotropic angular distributions could lead to situations with much larger instability growth rates~\cite{Sawyer:2005jk,Sawyer:2008zs}, a great number of studies focused on the ``fast" flavor instability (FFI)~\cite{Richers_review,WuTamborra,Abbar:2018shq,Morinaga:2019wsv,Abbar:2019zoq,Nagakura:2021hyb,Richers:2022dqa,Grohs:2022fyq,Richers:2022bkd,Froustey:2023skf,Mukhopadhyay:2024zzl,Xiong:2024tac,Xiong:2024pue,Fiorillo:2024bzm,Fiorillo:2024uki}. More recently, another regime of instabilities, driven by collisions, was uncovered by Johns~\cite{Johns:2021qby}. Despite the common expectation that collisions damp flavor coherence, the interplay between the self-interaction mean-field potential and a difference of interaction rates between neutrinos and antineutrinos can actually trigger a ``collisional" flavor instability (CFI). This initial discovery was followed by many works (e.g.,~\cite{Padilla-Gay:2022wck,Johns:2022yqy,Lin:2022dek,Xiong:2022zqz,Fiorillo:2023ajs,Liu:2023pjw}), with several focusing on the role of CFIs in supernovae~\cite{Xiong:2022vsy,Liu:2023vtz,Akaho:2023brj,Shalgar:2023aca,Zaizen:2024faj,Wang:2025vbx}.

To determine whether neutrino flavor transformation can occur for a given flavor configuration, it is natural to perform a linear stability analysis (LSA)~\cite{Banerjee:2011fj}. This analysis requires linearizing the quantum kinetic equations (QKEs), ideally written for all of the system's momentum degrees of freedom: the different propagation directions (angular distribution) and the magnitudes (energies). We base our LSA on the framework of Ref.~\cite{Froustey:2023skf}, which derived the LSA equations in the form of a stability matrix eigenvalue problem for a single-energy system and employed only the first few angular moments in place of the full angular distribution. We generalize this formalism by considering a multi-energy spectrum, and include the vacuum and collision terms from the QKEs. We frequently discuss the different unstable modes and their characteristics in light of the recent comprehensive study by Fiorillo and Raffelt~\cite{Fiorillo:2025zio}, which uses an equivalent formalism to the stability matrix method, the so-called ``dispersion relation approach"~\cite{Izaguirre:2016gsx}.

Some recent studies~\cite{Xiong:2022zqz,Froustey:2023skf,Nagakura:2025hss} have looked at the occurrence of CFI in classical postmerger simulation snapshots, and they all estimated the growth rate assuming that the neutrino distributions were isotropic, an assumption that fails beyond the neutrino decoupling region. Furthermore, it is common in these studies to invoke growth rate formulas for a monochromatic gas using energy-averaged collision rates, following analytic and numerical arguments in~\cite{Lin:2022dek,Liu:2023pjw}. It was, however, shown recently in~\cite{Wang:2025vbx} that using energy-averaged collision rates dramatically overestimates the occurrence of CFIs in CCSNe. In this paper we examine this same, energy-averaged, method of stability analysis using a NSM simulation snapshot from~\cite{Foucart:2024npn}, and find that the different properties of the flavor-dependent neutrino distributions in CCSNe and NSMs actually allow for a good energy-averaged description in the latter. Our work aims at providing a complete picture of the possible instabilities that can occur in a NSM postmerger environment, describing slow, collisional, and fast modes.

A LSA informs us of the locations and local properties (wavelength, timescale) of flavor instabilities, but not on their eventual, asymptotic, outcome. This aspect is beyond the scope of this paper but is being intensely investigated for fast instabilities~\cite{Bhattacharyya:2020jpj,Bhattacharyya:2022eed,Zaizen:2023ihz,Xiong:2023vcm,Richers:2024zit,George:2024zxz,Fiorillo:2024qbl,Liu:2024nku,Liu:2025tnf,Urquilla:2025idk}, which has allowed for the integration of subgrid models of the FFI in large-scale simulations of CCSNe~\cite{Wang:2025nii,Wang:2025ihh,Akaho:2026kff} and postmerger disks~\cite{Lund:2025jjo}. The outcome of the CFI has been less intensely studied; see, e.g.,~\cite{Zaizen:2025ptx,Froustey:2025nbi}.

Our paper is organized as follows. In Sec.~\ref{sec:evolution_equations}, we introduce the neutrino evolution equations and their version written for angular moments, which we linearize to derive our LSA framework. For the specific case of CFIs with isotropic neutrino distributions, we discuss various monochromatic descriptions of the system (see also Appendix~\ref{app:CFI_iso} for useful analytic formulas). We discuss in Appendix~\ref{app:dispersion_relation} the connection of our LSA method, based on a stability matrix, with the commonly used ``dispersion relation" approach. We introduce in Sec.~\ref{sec:methods} the NSM simulation data we study in this work, outlining the approximations we make to obtain the angular and energy neutrino distributions at each simulation grid point. We then describe the neutrino flavor instabilities occurring in this NSM snapshot, first assuming in Sec.~\ref{sec:results_isotropic_CFI}, for the sake of comparison with the CFI literature, that neutrino distributions are isotropic. We notably discuss how energy-averaged methods can provide a good estimate of the CFI growth rate in this environment, with a study of the difference with CCSN configurations in Appendix~\ref{app:compareWang}. We investigate the changes in the CFI landscape when including nucleon many-body corrections to the absorption rates (Sec.~\ref{subsec:manybody}), scattering opacities (Sec.~\ref{subsec:iso_scatt}), or the vacuum term (Sec.~\ref{subsec:iso_vac}). In Sec.~\ref{sec:results_aniso}, we study the flavor instabilities in the NSM snapshot when considering the actual neutrino anisotropies, which allows us to describe fast, slow, and collisional instabilities. Our findings are confirmed in other NSM snapshots, which we discuss in Appendix~\ref{app:other_snapshots}. Finally, we summarize and conclude in Sec.~\ref{sec:conclusion}.

Throughout this paper, we use natural units in which $\hbar = c = k_\mathrm{B} = 1$, and write $\hat{\vec{v}} = \vec{v}/\lVert \vec{v} \rVert$ for unit vectors.

%%%%%%%%=======================%%%%%%%%
\section{Neutrino evolution equations}
\label{sec:evolution_equations}
%%%%%%%%=======================%%%%%%%%

%%%%%%%%%%
\subsection{Quantum Kinetic Equations}
%%%%%%%%%%

The statistical ensemble of (anti)neutrinos, including flavor mixing, is described by one-body density matrices for neutrinos [$\vrho(t,\vec{x},\vec{p})$] and antineutrinos [$\bvrho(t,\vec{x},\vec{p})$], where $t$ is time, $\vec{x}$ is the spatial coordinates, and $\vec{p}$ is the (anti)neutrino momentum. They are matrices in flavor space, where the on-diagonal components generalize the classical distribution functions (we will sometimes use the notation $f_{\nu_\alpha} = \vrho_{\alpha \alpha}$), and the complex off-diagonal components measure the so-called ``flavor coherence." Their evolution is given by the QKEs~\cite{SiglRaffelt,Blaschke:2016xxt,Richers:2019grc,Froustey:2020mcq}
\begin{equation}
\label{eq:QKE}
\begin{aligned}
\ii \left( \pdv{\vrho_{\alpha \beta}}{t} + \dot{\vec{x}} \cdot \vec{\nabla} \vrho_{\alpha \beta} \right) &= \left[H, \vrho\right]_{\alpha \beta} + \ii \,  C_{\alpha \beta} \, ,\\
\ii \left( \pdv{\bvrho_{\alpha \beta}}{t} + \dot{\vec{x}} \cdot \vec{\nabla} \vrho_{\alpha \beta} \right) &= \left[\bar{H}, \bvrho\right]_{\alpha \beta} + \ii \,  \overline{C}_{\alpha \beta} \, ,
\end{aligned}
\end{equation}
where $H$ is the Hamiltonian operator including both vacuum and mean-field interaction contributions, $H = H_V + H_M + H_{\nu \nu}$. In this paper, we restrict to two-flavor systems, with the electron flavor (anti)neutrinos $\nu_e$ ($\bar{\nu}_e$) and the heavy lepton flavor (anti)neutrinos $\nu_x$ ($\bar{\nu}_x$). The vacuum Hamiltonian then reads in the flavor basis
\begin{equation}
    H_V = \frac{\Delta m^2}{4p} \begin{pmatrix} - \cos(2 \theta) & \sin(2\theta) \\ \sin(2 \theta) & \cos(2 \theta) \end{pmatrix} \, ,
\end{equation}
where $p=\lvert \vec{p} \rvert$. In this work, we take $\Delta m^2 = \Delta m^2_{31} = \pm 2.5 \times 10^{-3} \, \mathrm{eV}^2$ (for normal and inverted ordering) and $\sin^2 (\theta) = \sin^2 (\theta_{13}) = 2.16 \times 10^{-2}$~\cite{ParticleDataGroup:2024cfk}. If one considers that the only charged leptons in the background are electrons (no muons, although their presence could have some effects~\cite{Bollig:2017lki,Fischer:2020vie,Capozzi:2020syn,Loffredo:2022prq,Liu:2024wzd,Gieg:2024jxs,Ng:2024zve}), and that we are in the frame comoving with the fluid, the matter term reads in the flavor basis
\begin{equation}
    H_M = \sqrt{2} G_F n_e \begin{pmatrix} 1 & 0 \\ 0 & 0 \end{pmatrix} \, ,
\end{equation}
with $G_F \simeq 1.166 \times 10^{-5} \, \mathrm{GeV}^{-2}$ Fermi's constant, and $n_e$ the electron number density. Finally, the self-interaction mean-field potential $H_{\nu \nu}$ is
\begin{equation}
\label{eq:H_self}
H_{\nu \nu} = \frac{\sqrt{2} G_F}{(2 \pi)^3}  \int{\dd^3 \vec{q} \, (1- \hat{\vec{p}}\cdot \hat{\vec{q}})[\vrho(t,\vec{x},\vec{q}) - \bvrho^*(t,\vec{x},\vec{q})]} \, .
\end{equation}
For antineutrinos we have $\bar{H} = H_V - H_M - H_{\nu \nu}^*$. We postpone the presentation of the collision terms to the end of this section as they naturally connect to the moment equations that we now introduce.

%%%%%%%%%%%%
\subsection{Moment formalism}
%%%%%%%%%%%%

A common strategy for the neutrino transport in classical (i.e., without flavor mixing) hydrodynamic simulations consists of evolving only a small set of angular-integrated quantities instead of the full angular distributions~\cite{thorne1981relativistic,Shibata:2011kx,Foucart:2022bth}. Generalizing this approach to density matrices, we can define the first quantum moments
\begin{subequations}
\label{eq:moments_p}
\begin{align}
    N_{\alpha \beta}(t,\vec{x},p) &\equiv \frac{p^2}{(2 \pi)^3} \int{\dd \vec{\Omega} \, \vrho_{\alpha \beta}(t,\vec{x},\vec{p})} \, , \label{eq:moment_p_N} \\
    F_{\alpha \beta}^i(t,\vec{x},p) &\equiv \frac{p^2}{(2 \pi)^3}  \int{\dd \vec{\Omega} \, \frac{p^i}{p} \, \vrho_{\alpha \beta}(t,\vec{x},\vec{p})} \, , \label{eq:moment_p_F} \\
    P_{\alpha \beta}^{ij}(t,\vec{x},p) &\equiv \frac{p^2}{(2 \pi)^3}  \int{\dd \vec{\Omega} \, \frac{p^i p^j}{p^2} \, \vrho_{\alpha \beta}(t,\vec{x},\vec{p})} \, , \label{eq:moment_p_P}
\end{align}
\end{subequations}
which we will, respectively, call “number density,” “(number) flux density,” and “pressure tensor.” It should be noted that $P_{\alpha \beta}^{ij}$ does not have here the units of a pressure, but we still use this name for convenience. The energy-integrated moments are
\begin{subequations}
\label{eq:moments_cal}
\begin{align}
    \N_{\alpha \beta}(t,\vec{x}) &\equiv \int{\dd{p} \, N_{\alpha \beta}(t,\vec{x},p)} \, , \\
    \mathcal{J}_{\alpha \beta}(t,\vec{x}) &\equiv \int{\dd{p} \, p \, N_{\alpha \beta}(t,\vec{x},p)} \, , \\
    \F_{\alpha \beta}^i(t,\vec{x}) &\equiv \int{\dd{p} \, F_{\alpha \beta}^i(t,\vec{x},p)} \, .
\end{align}
\end{subequations}
These energy-integrated moments are the quantities appearing in the QKE Hamiltonian term in Eq.~\eqref{eq:H_self}, which can now be written as $H_{\nu \nu} = \sqrt{2} G_F [(\N - \overline{\N}^*) - \hat{\vec{p}} \cdot (\bm{\F} - \bm{\overline{\F}}^*)]$. We have also introduced the energy density $\mathcal{J}$, which will be a useful quantity in Sec.~\ref{subsec:data_NSM} to define energy spectra. In terms of moments, the QKEs become (we use Einstein's summation convention)
\begin{widetext}
\begin{subequations}
\begin{align}
\ii \left(\pdv{N}{t} + \pdv{F^j}{x^j} \right) &= \left[H_V + H_M,N\right] + \sqrt{2} G_F \left[\N - \overline{\N}^*, N\right] - \sqrt{2} G_F  \left[(\F-\overline{\F}^*)_j,F^j\right] + \ii \, C_N \, ,  \label{eq:QKE_moment_N} \\
\ii \left(\pdv{F^i}{t} + \pdv{P^{ij}}{x^j}\right) &= \left[H_V + H_M, F^i\right] + \sqrt{2} G_F \left[\N - \overline{\N}^*, F^i\right] - \sqrt{2} G_F \left[(\F-\overline{\F}^*)_j,P^{ij}\right] + \ii \, C_F^i \, . \label{eq:QKE_moment_F}
\end{align}
\end{subequations}
$C_N$ and $C_F^{i}$ are angular integrals of the collision term $C$ appearing in Eq.~\eqref{eq:QKE}, similarly to Eqs.~\eqref{eq:moment_p_N} and \eqref{eq:moment_p_F}. The same equations for the antineutrino moments are
\begin{subequations}
\begin{align}
\ii \left(\pdv{\bN}{t} + \pdv{\bF^j}{x^j} \right) &= \left[H_V - H_M,\bN\right] - \sqrt{2} G_F \left[\N^* - \overline{\N}, \bN \right] + \sqrt{2} G_F  \left[(\F^*-\overline{\F})_j,\bF^j\right] + \ii \, \bC_N \, ,  \label{eq:QKE_moment_Nbar} \\
\ii \left(\pdv{\bF^i}{t} + \pdv{\bP^{ij}}{x^j}\right) &= \left[H_V - H_M,\bF^i\right] - \sqrt{2} G_F \left[\N^* - \overline{\N}, \bF^i\right] + \sqrt{2} G_F \left[(\F^*-\overline{\F})_j,\bP^{ij}\right] + \ii \, \bC_F^i \, . \label{eq:QKE_moment_Fbar}
\end{align}
\end{subequations}
\end{widetext}

In order to obtain a genuine two-moment scheme, that is, evolving only $(N,\bN)$ and $(\vec{F},\vec{\bF})$, one needs to express the pressure tensors as a function of the first two moments. In particular, since flavor evolution can lead to a transfer of power to smaller angular scales~\cite{Johns:2019izj,Johns:2020qsk}, this emphasizes the need for a proper \emph{closure}. Although the closure problem for classical neutrino transport has been intensely studied (see, e.g.,~\cite{Smit_closure,Pons:2000br,LareckiBanach,Murchikova:2017zsy,Richers:2020ntq,Wang:2022voe}), the generalization to quantum moments has only been recently explored; see~\cite{Froustey:2024sgz,Kneller:2024buy}. In this work, we perform a linear stability analysis restricted to a specific wavenumber, for which the closure relation for the flavor off-diagonal components of the moments plays no role, and we leave a dedicated study of the impact and adequacy of the chosen quantum closure relation for future work.

%%%%%%%%%%%%%%%%%%%%%%%%%%%%
\subsection{Collision terms}
%%%%%%%%%%%%%%%%%%%%%%%%%%%%

In this work, we consider interactions of (anti)neutrinos with matter through emission/absorption and \emph{isotropic} scattering reactions. We adopt a “relaxation-time” approximation, such that the collision terms are written
\begin{equation}
\label{eq:coll_moments}
    \begin{aligned}
        C_N &= \frac12 \left\{ \begin{pmatrix} \kappa_{\mathrm{a},e} & 0 \\ 0 & \kappa_{\mathrm{a},x} \end{pmatrix} , N^\mathrm{(eq)} - N \right\} \, , \\
        C_F^i &= - \frac12 \left\{ \begin{pmatrix} \kappa_{\mathrm{a},e} + \kappa_{\mathrm{s},e} & 0 \\ 0 & \kappa_{\mathrm{a},x} + \kappa_{\mathrm{s},x} \end{pmatrix} , F^i \right\} \, ,
    \end{aligned}
\end{equation}
where $\kappa_{\mathrm{a},\alpha}$ (resp.~$\kappa_{\mathrm{s},\alpha}$) are the energy-dependent absorption (resp.~scattering) opacities for $\nu_\alpha$. $N^\mathrm{(eq)}$ is the number density moment of the classical equilibrium state, which is a diagonal matrix in flavor space such that the emissivities $\eta_{\alpha}$ and absorption opacities are related by $\eta_{\alpha} = \kappa_{\mathrm{a},\alpha} N_{\alpha \alpha}^\mathrm{(eq)}$. The flavor-diagonal components of Eq.~\eqref{eq:coll_moments} are consistent with the collision terms appearing in classical moment-based simulations of dense astrophysical environments, see for instance Eq.~(A32) in~\cite{Foucart:2016rxm} or Eq.~(8) in~\cite{Foucart:2017mbt}. Considering the flavor off-diagonal components, we see that the damping rates of flavor coherence are
\begin{equation}
\label{eq:collision_N_F}
    \begin{aligned}
        - (C_N)_{ex} &\equiv \Gamma_N N_{ex} = \frac{\kappa_{\mathrm{a},e} + \kappa_{\mathrm{a}, x}}{2} N_{ex} \, , \\
        - (C_F)^i_{ex} &\equiv \Gamma_F F_{ex}^i = \left(\frac{\kappa_{\mathrm{a}, e} + \kappa_{\mathrm{a}, x}}{2} + \frac{\kappa_{\mathrm{s}, e} + \kappa_{\mathrm{s}, x}}{2}\right) F_{ex}^i \, .
    \end{aligned}
\end{equation}
In the following, when the distinction between $\Gamma_N$ and $\Gamma_F$ is not relevant, we may simply write $\Gamma$ for brevity. While our use of these collision terms is an approximation of the whole set of processes that should ideally be included and may modify the CFI (see~\cite{Richers:2019grc} for a general discussion of the QKE collision term in astrophysical environments), we use this approximation for consistency with the NSM simulation we will study.

%%%%%%%%%%%%%%%%%%%%%%%%%%%%%%%%%%%%%%%
\subsection{Linear stability analysis}
\label{subsec:LSA}
%%%%%%%%%%%%%%%%%%%%%%%%%%%%%%%%%%%%%%%

We determine the occurrence of flavor instabilities by performing a linear stability analysis (LSA) of the QKEs—see~\cite{Froustey:2023skf} for the same moment-LSA approach in the monochromatic case. Specifically, we expand $N_{ex}(t,\vec{x},p) = A_{ex}(p) e^{- \ii (\Omega t - \vec{k} \cdot \vec{x})}$, $F_{ex}^i(t,\vec{x},p) = B_{ex}^i(p) e^{-\ii(\Omega t - \vec{k}\cdot \vec{x})}$, and likewise for antineutrinos, with constant populations of the flavor on-diagonal moments $N_{ee}(p)$, $N_{xx}(p)$, etc. Furthermore, the energy spectrum is binned into $n_\mathrm{groups}$ energy groups, centered on the energies $p_n$ and of width $\Delta p_n$. In the following, we write $A_{ex}(p_n) = A_{ex}^{(n)}$ and similarly for all energy-dependent quantities.

The linearized QKEs naturally involve the “shifted” frequency and wavenumber
\begin{equation}
\label{eq:omegaprime_kprime}
    \begin{aligned}
        \Omega' &\equiv \Omega - \sqrt{2} G_F n_e - \sqrt{2} G_F \left[\N_{ee} - \N_{xx} - \overline{\N}_{ee} + \overline{\N}_{xx}\right] \, , \\
        \vec{k}' &\equiv \vec{k} - \sqrt{2} G_F \left[\bm{\F}_{ee} - \bm{\F}_{xx} - \bm{\overline{\F}}_{ee} + \bm{\overline{\F}}_{xx}\right] \, .
    \end{aligned}
\end{equation}
In the rest of this work, we restrict the analysis to the “zero-mode” $\vec{k}' = \vec{0}$. For this specific choice of wavevector, the linearized QKEs do not involve the flavor off-diagonal components of the pressure tensor, such that we do not need to specify a fully quantum closure~\cite{Froustey:2024sgz,Kneller:2024buy}. Although focusing on this zero-mode may miss some instability regions (see e.g.,~\cite{Dasgupta:2018ulw,Richers:2022dqa,Froustey:2023skf} for the FFI case), it provides a conservative estimate free of spurious modes. We also note that in the isotropic limit, it was shown in~\cite{Liu:2023pjw} that the zero-mode (which is then the homogeneous mode) has the largest CFI growth rate. 

The linearized QKEs then read\footnote{Technically, the linearized version of the $ex$ component of Eq.~\eqref{eq:QKE_moment_N} involves, on the right-hand side, a source term $ -\Delta m^2/(4p) \sin(2 \theta) [N_{ee} - N_{xx}]$, which is not proportional to $e^{- \ii (\Omega t - \vec{k} \cdot \vec{x})}$. Starting from pure flavor states, this term seeds flavor coherence. The (in)stability question is answered by looking at the homogeneous part of the equation, which admits or not exponentially growing solutions: these are the equations we report in Eq.~\eqref{eq:linearized_QKEs}. This point is commented upon in, e.g.,~\cite{Fiorillo:2024pns,Fiorillo:2025zio}, see also the derivation in~\cite{Lin:2022dek} where their matrix $\Lambda$ corresponds to our stability matrix $\mathsf{S}$.}

\begin{widetext}
\begin{subequations}
\label{eq:linearized_QKEs}
\begin{align}
\Omega' A_{ex}^{(n)} &= - \omega^{(n)} A_{ex}^{(n)} - \ii \, \Gamma_{N}^{(n)} A_{ex}^{(n)} - \Delta_N^{(n)} \sum_{m} \Delta p_m \left[A_{ex}^{(m)} - \bA_{xe}^{(m)}\right]  + \Delta_F^{j,(n)} \sum_{m} \Delta p_m \left[B_{ex,j}^{(m)}-\bB_{xe,j}^{(m)}\right] \, ,
\label{eq:LSA_N} \\
\Omega' B_{ex}^{i,(n)} &= - \omega^{(n)} B_{ex}^{i,(n)} - \ii \, \Gamma_F^{(n)} B_{ex}^{i,(n)} -\Delta_{F}^{i,(n)} \sum_{m} \Delta p_m \left[A_{ex}^{(m)} - \bA_{xe}^{(m)}\right] + \Delta_P^{ij,(n)} \sum_{m} \Delta p_m \left[B_{ex,j}^{(m)} - \bB_{xe,j}^{(m)}\right] \, ,
 \label{eq:LSA_F} \\
\Omega' \bA_{xe}^{(n)} &= + \omega^{(n)} \bA_{xe}^{(n)} - \ii \, \bGamma_{N}^{(n)} \bA_{xe}^{(n)} - \bDelta_N^{(n)} \sum_{m} \Delta p_m \left[A_{ex}^{(m)} - \bA_{xe}^{(m)}\right]  + \bDelta_F^{j, (n)} \sum_{m} \Delta p_m \left[B_{ex,j}^{(m)}-\bB_{xe,j}^{(m)}\right] \, ,
\label{eq:LSA_Nbar} \\
\Omega' \bB_{xe}^{i,(n)} &= + \omega^{(n)} \bB_{xe}^{i,(n)} - \ii \bGamma_F^{(n)} \bB_{xe}^{i,(n)} - \bDelta_F^{i, (n)} \sum_{m} \Delta p_m \left[A_{ex}^{(m)} - \bA_{xe}^{(m)}\right] + \bDelta_P^{ij,(n)} \sum_{m} \Delta p_m \left[B_{ex,j}^{(m)} - \bB_{xe,j}^{(m)}\right] \, . \label{eq:LSA_Fbar}
\end{align}
\end{subequations}
\end{widetext}
We introduced the vacuum Hamiltonian strength $\omega^{(n)} \equiv \Delta m^2 / (2p_n) \cos(2\theta)$, and the shorthand notations %$\Delta_N^{(n)} \equiv \sqrt{2} G_F [N_{ee}^{(n)} - N_{xx}^{(n)}]$, $\Delta_F^{j,(n)} \equiv \sqrt{2} G_F [F_{ee}^{j,(n)} - F_{xx}^{j,(n)}]$, and $\Delta_P^{ij,(n)} \equiv \sqrt{2} G_F [P_{ee}^{ij,(n)} - P_{xx}^{ij,(n)}]$.

\begin{equation}
    \begin{aligned}
        \Delta_N^{(n)} &\equiv \sqrt{2} G_F \left[N_{ee}^{(n)} - N_{xx}^{(n)}\right] \, , \\ 
        \Delta_F^{j,(n)} &\equiv \sqrt{2} G_F \left[F_{ee}^{j,(n)} - F_{xx}^{j,(n)}\right] \, , \\
        \Delta_P^{ij,(n)} &\equiv \sqrt{2} G_F \left[P_{ee}^{ij,(n)} - P_{xx}^{ij,(n)}\right] \, .
    \end{aligned}
\end{equation}

We turn the right-hand side of Eq.~\eqref{eq:linearized_QKEs} into a matrix equation, such that determining $\Omega'$ amounts to solving the eigenvalue problem
\begin{equation}
\label{eq:stability_matrix}
    \mathsf{S} \begin{bmatrix} \vdots \\ A_{ex}^{(n)} \\ B_{ex,\x}^{(n)} \\ B_{ex,\y}^{(n)} \\ B_{ex,\z}^{(n)} \\ \vdots \\ \bA_{xe}^{(m)} \\ \bB_{xe,\x}^{(m)} \\ \bB_{xe,\y}^{(m)} \\ \bB_{xe,\z}^{(m)} \\ \vdots \end{bmatrix} = \Omega' \begin{bmatrix} \vdots \\ A_{ex}^{(n)} \\ B_{ex,\x}^{(n)} \\ B_{ex,\y}^{(n)} \\ B_{ex,\z}^{(n)} \\ \vdots \\ \bA_{xe}^{(m)} \\ \bB_{xe,\x}^{(m)} \\ \bB_{xe,\y}^{(m)} \\ \bB_{xe,\z}^{(m)} \\ \vdots \end{bmatrix} \, ,
\end{equation}
with $\mathsf{S}$ the stability matrix, whose components can be read from Eqs.~\eqref{eq:LSA_N}--\eqref{eq:LSA_Fbar}. Among the $8 \times n_\mathrm{groups}$ eigenvalues $\Omega'_s$ of $\mathsf{S}$, we define $\Omega'_\mathrm{max}$ as the one with the largest imaginary part, such that
\begin{equation}
    \Im(\Omega'_\mathrm{max}) = \max_{s} \left\{\Im(\Omega'_s)\right\} \, ,
\end{equation}
if positive, is the growth rate of the instability.

We emphasize that the specific choice $\vec{k}' = \vec{0}$ reduces the stability matrix dependence on the pressure tensors to their flavor on-diagonal components, which appear in $\Delta_P^{ij}$, $\bDelta_P^{ij}$. For other values of $\vec{k}'$, one needs to also specify the closure for the off-diagonal component $P_{ex}^{ij}$; see Ref.~\cite{Froustey:2023skf}. The NSM simulation we study in this work uses the classical M1 two-moment scheme, such that we can borrow its closure prescription for $P_{\alpha \alpha}^{ij}$ [see Eq.~\eqref{eq:P_class}].

\subsection{Energy-averaged approaches}
\label{subsec:energy_averaging}

Much of the work on CFIs has focused on monochromatic energy distributions, or at least used the results of the monochromatic dispersion relation with certain energy-averaged quantities (e.g.,~\cite{Johns:2021qby,Johns:2022yqy,Liu:2023vtz,Akaho:2023brj,Liu:2024wzd,Froustey:2025nbi}). It is an attractive approach as it avoids the numerical complexity of the multi-energy analysis: for instance, it reduces by a factor $n_\mathrm{groups}^2$ the size of the stability matrix [Eq.~\eqref{eq:stability_matrix}]. Such methods were justified by analytical and numerical results~\cite{Lin:2022dek,Xiong:2022zqz,Liu:2023pjw}, which showed that the instability growth rate in multi-energy analyses was well recovered by monochromatic formulas.

Recently, in their work assessing the impact of CFIs on core-collapse supernovae, Wang \emph{et al.}~\cite{Wang:2025vbx} have pointed out some strong limitations of these energy-averaging justifications. In particular, the regime in which a monochromatic effective treatment would be possible is not the one found in CCSNe, which has led various works to overestimate the occurrence of CFIs in CCSNe. In their comprehensive study of the solutions of the dispersion relation describing neutrino flavor instabilities, Fiorillo and Raffelt~\cite{Fiorillo:2025zio} make a similar argument, emphasizing the difference between ``gapped" modes (for which an analytical reduction to an effective monochromatic system is possible, as done in~\cite{Lin:2022dek,Xiong:2022zqz}), and ``gapless" modes, which would be prevalent in CCSNe and for which such a matching is not justified. In this section, we discuss the two previously used energy-averaging methods and the analytic justification for the first, as they will be compared with multi-energy results in the following.

An important result of this work, that we detail in Sec.~\ref{sec:results_isotropic_CFI}, is that collisional instabilities of both the gapless and the gapped kind appear in NSM environments. Furthermore, we show that it is possible to approximately describe the instabilities of the latter type with energy-averaged quantities, up to a rescaling. For these reasons, we present in Appendix~\ref{app:CFI_iso} the single-energy formulas previously obtained in the literature, but derived from our moment approach---see in particular Eqs.~\eqref{eq:disp_isopres} and \eqref{eq:disp_isobreak}. Then, in Appendix~\ref{app:dispersion_relation}, we show the equivalence between our stability matrix method and the commonly used dispersion relation approach~\cite{Izaguirre:2016gsx,Fiorillo:2025zio}.

Before we proceed, let us summarize and clarify the nomenclature used to describe the CFI modes. If one assumes that the flavor-diagonal distributions are isotropic, the structure of the stability matrix simplifies and we can identify different types of modes (see Appendix~\ref{subsec:isobreak_pres}): the “isotropy-preserving" ones, which are pure number density perturbations ($B_{ex,j}^{(n)}=0$), and the “isotropy-breaking" ones, which are pure flux perturbations ($A_{ex}^{(n)} = 0$). For the latter, although the flavor on-diagonal fluxes vanish (isotropy assumption), the flavor coherence wave develops a nonzero flux—hence the “breaking" of isotropy. When the flavor on-diagonal fluxes do not vanish, this classification no longer holds and all modes have both a number and a flux component. Separate from this classification, we can also distinguish the eigenmodes based on the real part of $\Omega'$. We discuss this point in Appendix~\ref{app:gapped_gapless}, but for all practical purposes the gapped (resp.~gapless) modes are characterized by $|\Re(\Omega')| \gg \Gamma$ (resp.~$|\Re(\Omega')| \simeq 0$). If one assumes isotropy, there are both gapless and gapped isotropy-preserving modes, and likewise for isotropy-breaking modes, but the gapless/gapped distinction extends beyond this limiting case. Finally, in the monochromatic limit, the gapped and gapless modes have also been called “minus" and “plus" modes. We make the explicit connection in Appendix~\ref{app:gapped_gapless}, but hereafter we will stick to the gapless/gapped nomenclature of~\cite{Fiorillo:2025zio} since it has a physical foundation and is not limited to monochromatic spectra.

\subsubsection{Gapped modes and energy-averaging ``method A"}

Assuming homogeneity and isotropy, let us consider a gapped mode for which $|\Re(\Omega')| \sim \sqrt{2} G_F |\N_{ee} - \N_{xx} - \overline{\N}_{ee} + \overline{\N}_{xx}| \gg \Gamma(p), \bGamma(p)$. Indeed, in a typical NSM environment, neutrino densities are such that $\sqrt{2} G_F \N \sim 10^{10}-10^{11} \, \mathrm{s}^{-1}$, while $\Gamma < 10^6 \, \mathrm{s}^{-1}$. Note that we omitted the subscript ``$N$'' on $\Gamma$ for brevity. Transforming Eq.~\eqref{eq:stability_matrix} as in Appendix~\ref{app:dispersion_relation} allows us to find the dispersion relation~\eqref{eq:dispersion_CFI}. Expanding it at first order in $|\Gamma / \Omega'|$, as in~\cite{Xiong:2022zqz}, produces
\begin{multline}
\label{eq:expansion_gapped}
    -1 \simeq \frac{1}{\Omega'} \sqrt{2} G_F \int_{0}^{\infty}{\frac{p^2 \dd{p}}{2 \pi^2}\left[f_{\nu_e} - f_{\nu_x} - f_{\bar{\nu}_e} + f_{\bar{\nu}_x} \right]} \\
    - \frac{\ii}{\Omega^{\prime \, 2}} \sqrt{2} G_F \int_{0}^{\infty}{\frac{p^2 \dd{p}}{2 \pi^2}\left[\left(f_{\nu_e} - f_{\nu_x}\right) \Gamma - \left(f_{\bar{\nu}_e} - f_{\bar{\nu}_x}\right) \bGamma\right]} \, .
\end{multline}
Multiplying \eqref{eq:expansion_gapped} through by $\Omega'^2$, we arrive at the expression $\Omega^{\prime \, 2} + D_0 \Omega' - \ii \Sigma_0 = 0$ (where we borrow notations from~\cite{Fiorillo:2025zio}), with solutions
\begin{equation}
    \Omega' = - \frac{D_0}{2} \pm \sqrt{\left(\frac{D_0}{2}\right)^2 + \ii \, \Sigma_0} \, .
\end{equation}
In the physical regime $|\Sigma_0| \ll |D_0|$, we obtain the two solutions\footnote{The expression for $\Omega'_-$ matches Eq.~(4.8) in~\cite{Fiorillo:2025zio}, up to a sign reversal of the real part, because their equations are written for the $xe$ component of the density matrix.}
\begin{equation}
\label{eq:multi_gapped}
\begin{aligned}
    \Omega'_- &\simeq - D_0 - \ii \frac{\Sigma_0}{D_0} \, , \\
    \Omega'_+ &\simeq \ii \frac{\Sigma_0}{D_0} \, ,
\end{aligned}
\end{equation}
where, as noted in~\cite{Wang:2025vbx}, only the first solution is consistent with the initial assumption $|\Re(\Omega')| \gg \Gamma$. Crucially, this first solution exactly matches the monochromatic result~\eqref{eq:minus_mode}, if one replaces $\Gamma$ with an average of collision rates over the isospin distribution:
\begin{equation}
    \label{eq:methodA}
    \langle \Gamma \rangle_\mathrm{A} \equiv \frac{\displaystyle\int_{0}^{\infty}{p^2 \dd{p} \left(f_{\nu_e} - f_{\nu_x}\right) \Gamma(p)}}{\displaystyle\int_{0}^{\infty}{p^2 \dd{p} \left(f_{\nu_e} - f_{\nu_x}\right)}} \, ,
\end{equation}
such that $D_0 = \sqrt{2} G_F \left(\N_{ee} - \N_{xx} - \overline{\N}_{ee} + \overline{\N}_{xx}\right) = \Delta_\N - \bDelta_\N$, and $\Sigma_0 = \langle \Gamma \rangle_\mathrm{A} \Delta_\N - \langle \bGamma \rangle_\mathrm{A} \bDelta_\N$.

One can therefore estimate the multi-energy isotropic CFI growth rate by using the full monochromatic formulas~\eqref{eq:disp_isopres} with the averaged collision rates~\eqref{eq:methodA}. This energy-averaging approach is called ``Method A" in~\cite{Wang:2025vbx}, and we adopt this nomenclature. Note that, as demonstrated above, the gapped modes should be correctly described by this method, but nothing guarantees that the gapless modes would be well recovered. Indeed, Wang \emph{et al.} have shown that the growth rates of gapless modes are significantly overestimated, even predicting instability when the system is actually stable. We return to this point in Sec.~\ref{sec:results_isotropic_CFI}.

\subsubsection{Energy-averaging ``method B"}

Another monochromatic description, used for instance in~\cite{Liu:2023pjw,Akaho:2023brj}, consists of averaging the flavor-dependent opacities $\kappa_{\alpha}$ (we omit the ``absorption" subscript) and then reconstructing the effective flavor coherence damping rate. Specifically, we define
\begin{equation}
\label{eq:gray_opacity}
    \langle \kappa_\alpha \rangle \equiv  \frac{\displaystyle \int_{0}^{\infty}{p^2 \dd{p} \, f_{\nu_\alpha}(p) \, \kappa_\alpha(p)}}{\displaystyle \int_{0}^{\infty}{p^2 \dd{p}\, f_{\nu_\alpha}(p)}} \, ,
\end{equation}
from which we build
\begin{equation}
    \label{eq:methodB}
    \langle \Gamma \rangle_\mathrm{B} \equiv \frac{\langle \kappa_e \rangle + \langle \kappa_x \rangle}{2} \, ,
\end{equation}
and use the monochromatic formula~\eqref{eq:disp_isopres}. This approach is dubbed ``Method B" by Wang \emph{et al.}~\cite{Wang:2025vbx}. It is appealing because the averaged opacities~\eqref{eq:gray_opacity} are similar to the ones used in gray neutrino transport schemes~\cite{Foucart:2022bth} (although those schemes are usually implemented at the level of energy density moments, resulting in an extra factor of $p$ in the integrals above). However, there is no \emph{a priori} analytical justification for this method, either for gapped or gapless modes.

%%%%%%%%%%%%%%%%%%%
\section{Input data and methods}
\label{sec:methods}
%%%%%%%%%%%%%%%%%%%

\begin{figure*}[!ht]
    \centering
    \includegraphics[width=0.96\textwidth]{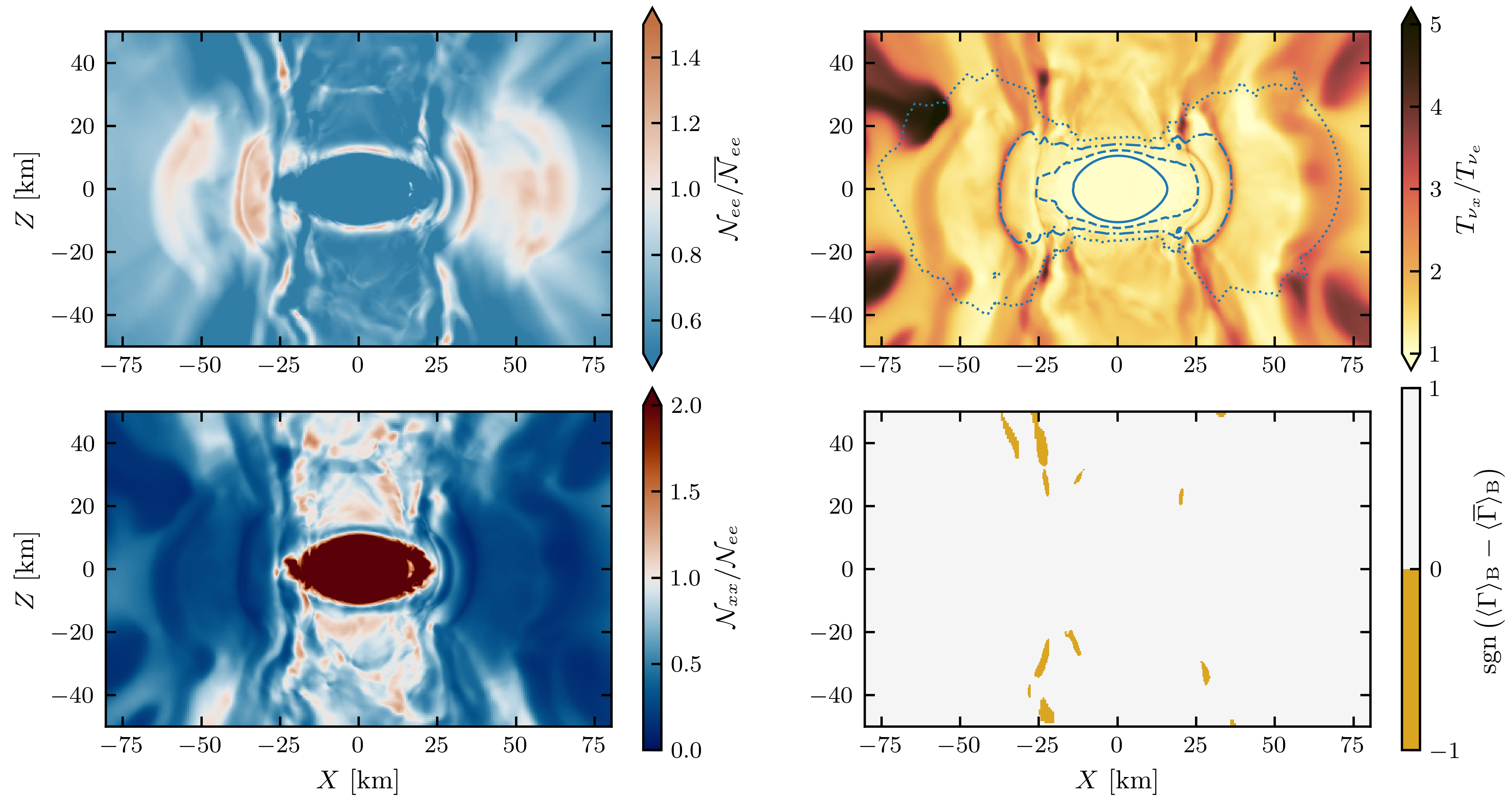}
    \caption{Data from the transverse slice at $Y = 0$ in the $7 \, \mathrm{ms}$ postmerger snapshot of the ``M1-NuLib" simulation in~\cite{Foucart:2024npn}. \emph{Top left:} ratio of $\nu_e$ and $\bar{\nu}_e$ number densities. \emph{Bottom left:} ratio of $\nu_x$ and $\nu_e$ number densities. \emph{Top right:} ratio of the effective temperatures (i.e., average energies) of $\nu_x$ and $\nu_e$, with matter density contours shown in blue (from the innermost to the outermost: $\{10^{14},10^{12},10^{11},10^{10}\} \, \mathrm{g \, cm^{-3}}$). \emph{Bottom right:} comparison of the energy-averaged absorption rates defined in Eq.~\eqref{eq:methodB}, showing the higher interaction rates of neutrinos over antineutrinos.}
    \label{fig:input_data}
\end{figure*}

%%%%%%%
\subsection{NSM simulation snapshots}
\label{subsec:data_NSM}
%%%%%%%

We apply our linear stability analysis to the results from a general relativistic simulation of the merger of two neutron stars ($1.3 \, M_\odot$ and $1.4 \, M_\odot$) which uses a gray two-moment scheme for neutrino transport~\cite{Foucart:2015vpa,Foucart:2016rxm}, specifically the “M1-NuLib” simulation in \cite{Foucart:2024npn}. We examine two different stages of the postmerger evolution, selecting snapshots at $3 \, \mathrm{ms}$ (“early”) and $7 \, \mathrm{ms}$ (“late”) post-merger---in the simulation, the hypermassive neutron star (HMNS) remnant collapses into a black hole after $8.5 \, \mathrm{ms}$. In the main text, we focus on the $7 \, \mathrm{ms}$ snapshot, with results for the $3 \, \mathrm{ms}$ snapshot shown in Appendix~\ref{app:other_snapshots} (along with a snapshot from an independent simulation, taken from Ref.~\cite{Foucart:2016rxm}). We show in Fig.~\ref{fig:input_data} some data from the simulation, for a vertical slice passing through the ``late'' snapshot of the remnant. The top-right panel of Fig.~\ref{fig:input_data} shows density contours for this $7 \, \mathrm{ms}$ postmerger snapshot. At that time, the HMNS is surrounded by an accretion torus of density $10^{10-12}\,{\rm g \, cm^{-3}}$. Within the torus are shocked tidal arms where the fluid temperature is higher and large density jumps can be seen. The tidal arms create sharp features in many of the vertical slices shown here (e.g., the region of high temperature $T_{\nu_x}$ at the density contour $\rho = 10^{11}\,{\rm g \, cm^{-3}}$ on Fig.~\ref{fig:input_data}). The polar regions above and below the remnant are filled with low-density outflowing matter.

Consistent with this simulation, we compute the collision rates from the neutrino interaction library \texttt{NuLib},\footnote{\texttt{NuLib} is available at \href{http://www.nulib.org}{http://www.nulib.org}.} using the temperature, matter density, average neutrino energy and electron fraction from the simulation snapshots and assuming the SFHo equation of state~\cite{Steiner:2012rk}. Specifically, we include the same microphysics for the neutrino interactions as in Ref.~\cite{Foucart:2024npn}: absorption on nucleons; isotropic and elastic scattering on nucleons, $\alpha$-particles and heavy nuclei; and electron-positron annihilations and nucleon-nucleon Bremsstrahlung for heavy-lepton (anti)neutrinos only. The table is logarithmically spaced in neutrino energies (16 groups\footnote{We checked that when using twice as many energy groups, the instability growth rates we obtain are not modified beyond the percent level. We thus stick to this table to be as close as possible to the assumptions of~\cite{Foucart:2024npn}.} up to $528 \, \mathrm{MeV}$), matter density $\rho$ (82 points from $10^6$ to $3.2 \times 10^{15} \, \mathrm{g \, cm^{-3}}$) and fluid temperature $T$ (65 points from $0.05$ to $150 \, \mathrm{MeV}$), and linearly spaced in the electron fraction $Y_e$ (51 points from $0.01$ to $0.6$).

The simulation snapshots provide the classical moments in a box of size $(X,Y,Z) \in [-160 \, \mathrm{km}, + 160 \, \mathrm{km}]$, divided in four refinement levels (cubes of side divided by two between adjacent levels). Each refinement level consists of a grid of $256^3$ points, and we do not consider data in any region of a refinement level that is covered by a finer refinement level.

\subsection{Neutrino distributions}
\label{subsec:neutrino_distrib}

The NSM simulation from which we extract the thermodynamics and neutrino data uses an energy-integrated two-moment scheme, from which we obtain the energy flux, energy density and number density in the inertial frame. Transforming these radiation field quantities into a frame comoving with the fluid (see, e.g., Appendix B in~\cite{Grohs:2023pgq}), we can get the quantities $\N_{\alpha \alpha}$ and $\F_{\alpha \alpha}^i$ [see definitions in Eq.~\eqref{eq:moments_cal}]. To perform our linear stability analysis, we need the energy distributions $N_{\alpha \alpha}(p)$ and $F_{\alpha \alpha}^i(p)$, and the flavor-diagonal pressure tensor $P_{\alpha \alpha}^{ij}$. In addition, this simulation considers three neutrino species: $\nu_e$, $\bar{\nu}_e$ and a combination of the heavy-lepton flavor neutrinos and antineutrinos, assumed to have the same distributions, $\nu_X$. In our two-flavor analysis, we thus take the values for $\nu_e$ and $\bar{\nu}_e$ and we assign schematically the distribution of a \textit{single} heavy-lepton neutrino species $f_{\nu_x} = f_{\nu_X}/4$, $f_{\bar{\nu}_x} = f_{\nu_X}/4$. In the left panels of Fig.~\ref{fig:input_data}, we show ratios of the neutrino number densities $\N_{\alpha \alpha}$, comparing $\nu_e$ and $\bar{\nu}_e$ in the top panel and $\nu_x$ and $\nu_e$ in the bottom panel.

We make the assumption that the classical neutrino distributions can be separated into an angular part and a spectral part, namely,\footnote{This assumption does not take into account the faster diffusion of low-energy neutrinos, which should lead technically to different spectral shapes for the energy and flux densities (see for instance the discussion in~\cite{Foucart:2016rxm}).}
\begin{equation}
\label{eq:distrib}
    f_{\nu_\alpha}(\vec{p}) = f_{\nu_\alpha}^\mathrm{ME}(\hat{\vec{p}}) \times f_{\nu_\alpha}^\mathrm{gray}(p) \, .
\end{equation}
We recall that we use a hat over a vector to indicate the associated unit vector. The angular part of $f_{\nu_\alpha}$ is given by the classical maximum entropy (ME), or Minerbo, closure~\cite{minerbo_maximum_1978,Smit_closure,Murchikova:2017zsy}:
\begin{equation}
\label{eq:distrib_ME}
    f_{\nu_\alpha}^\mathrm{ME}(\hat{\vec{p}}) = \frac{1}{4 \pi} \frac{Z_\alpha}{\sinh(Z_\alpha)} e^{Z_\alpha \hat{\vec{p}} \cdot \widehat{\bm{\F}}_{\alpha \alpha}} \, ,
\end{equation}
with
\begin{equation}
\label{eq:Z_ME}
    \coth(Z_\alpha) - \frac{1}{Z_\alpha} = \frac{\lVert \bm{\F}_{\alpha \alpha} \rVert}{\N_{\alpha \alpha}} \equiv \tilde{f}_\alpha \, .
\end{equation}
With this choice of closure, the pressure moment in the fluid frame is expressed as an interpolation between the optically thick and thin limits
\begin{equation}
\label{eq:P_class}
    P^{ij}_{\alpha \alpha} = \frac{3(1-\chi_\alpha)}{2} \frac{N_{\alpha \alpha}}{3} \delta^{ij} + \frac{3 \chi_\alpha -1}{2} \frac{F^i_{\alpha \alpha} F^j_{\alpha \alpha}}{|\vec{F}_{\alpha \alpha}|^2} N_{\alpha \alpha}  \, ,
\end{equation}
where the Eddington factor $\chi_\alpha$ is given as a function of the flux factor $\tilde{f}_\alpha$ by~\cite{cernohorsky_bludman}
\begin{equation}
\label{eq:chi_MEC}
  \chi_\alpha = 1 - \frac{2 \tilde{f}_\alpha}{Z_\alpha} \simeq \frac{1}{3} + \frac{2 \tilde{f}_\alpha^2}{15}\left(3-\tilde{f}_\alpha+3 \tilde{f}_\alpha^2\right)\, .
\end{equation}

The energy spectrum is assumed to be a gray Fermi-Dirac distribution:
\begin{equation}
\label{eq:distrib_gray}
    f_{\nu_\alpha}^\mathrm{gray}(p) = \frac{g_{\nu_\alpha}}{e^{p/T_{\nu_\alpha}} + 1} \, , 
\end{equation}
where the parameters $\{g_{\nu_\alpha}, T_{\nu_\alpha}\}$ are determined by the two equations:
\begin{equation}
    \begin{aligned}
        \frac{1}{(2 \pi)^3} \int_{0}^{\infty}{p^2 \dd{p} \, f_{\nu_\alpha}^\mathrm{gray}(p)} &= \N_{\alpha \alpha} \, , \\
        \frac{1}{(2 \pi)^3} \int_{0}^{\infty}{p^3 \dd{p} \, f_{\nu_\alpha}^\mathrm{gray}(p)} &= \mathcal{J}_{\alpha \alpha} \, .
    \end{aligned}
\end{equation}
Note that with this choice, the ``temperature" corresponds to the average energy $T_{\nu_\alpha}=\mathcal{J}_{\alpha \alpha}/\N_{\alpha \alpha}$. We have verified that using a Fermi-Dirac spectrum with a chemical potential instead of the gray distribution~\eqref{eq:distrib_gray} leads to the same regions of instability and very similar growth rates. We thus adopt the gray distribution for its simplicity.

In the top right panel of Fig.~\ref{fig:input_data}, we show the ratio of the temperatures $T_{\nu_\alpha}$ entering the gray spectrum~\eqref{eq:distrib_gray} for $\alpha=e,x$. We also show some matter density contours (see caption and discussion at the beginning of Sec.~\ref{sec:methods}). Finally, in the bottom right panel is shown which of the neutrinos or antineutrinos have the largest collision rate (restricting to absorption processes), focusing on the flavor-averaged collision rates~\eqref{eq:methodB}. In the large majority of the slice, neutrinos have larger interaction rates, such that the gapless or gapped nature of CFIs is determined by which of $\nu_e$ or $\bar{\nu}_e$ is the most abundant (see top left panel).

%%%%%%%%%%%%%%%%%%%
\section{Results for isotropic neutrino distributions}
\label{sec:results_isotropic_CFI}
%%%%%%%%%%%%%%%%%%%

In this section, we apply our linear stability analysis to the NSM simulation snapshot introduced in Sec.~\ref{subsec:data_NSM}. However, we consider that the classical neutrino distributions are \emph{isotropic}, such that the flavor-diagonal moments satisfy 
\begin{equation}
\label{eq:isotropic_background}
    \vec{F}_{\alpha \alpha} = \vec{0} \, , \qquad P^{ij}_{\alpha \alpha} = \frac{N_{\alpha \alpha}}{3} \delta^{ij} \, ,
\end{equation}
and likewise for antineutrinos. Since the flavor-diagonal fluxes vanish, $\vec{k'} = \vec{k}$ per Eq.~\eqref{eq:omegaprime_kprime}, such that we look for \emph{homogeneous} modes $\vec{k} = \vec{0}$. This isotropy assumption is clearly broken if we look further away from the HMNS, where neutrinos are free-streaming, but this method has been commonly used in the literature to assess the prevalence of CFIs. Note in particular that assuming isotropy prevents the appearance of fast instabilities. For consistency with this isotropy assumption and previous literature, we only include absorption processes here, such that $\Gamma_N = \Gamma_F$. We will look specifically at the changes incurring when one includes scattering opacities, still assuming that the neutrino flavor on-diagonal distributions are isotropic, in Sec.~\ref{subsec:iso_scatt}). A complete treatment with anisotropic neutrino distributions is postponed to Sec.~\ref{sec:results_aniso}.

\subsection{Isotropic CFI}
\label{subsec:isotropic_CFI}

\subsubsection{Multi-energy results}

\begin{figure}[!ht]
    \centering
    \includegraphics[width=\columnwidth]{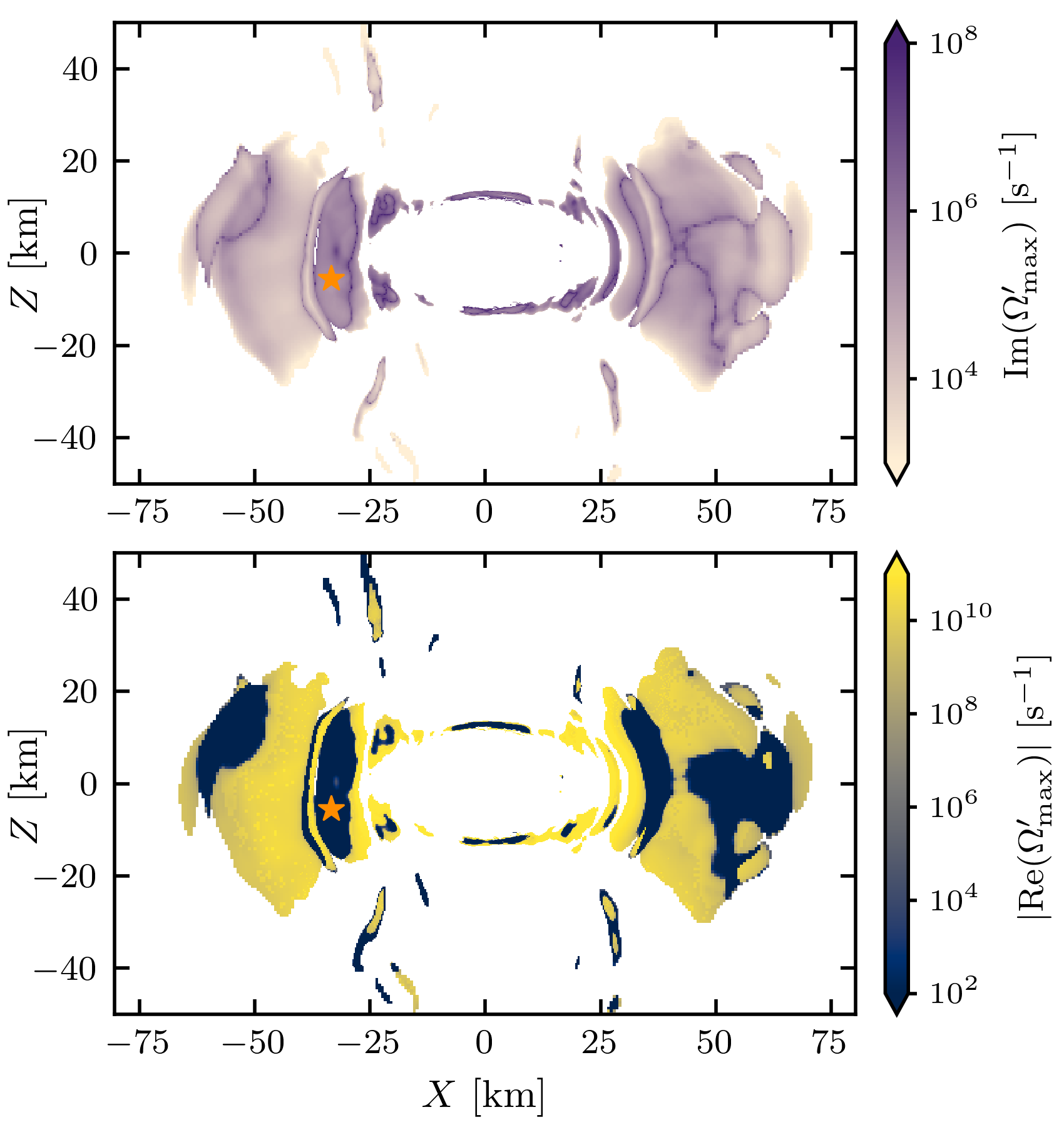}
    \caption{Result of the multi-energy homogeneous and isotropic LSA, without the vacuum term in the QKEs. \emph{Top:} growth rate of the instability. \emph{Bottom:} real part of the eigenfrequency of the fastest growing mode, enabling the identification of regions of gapless and gapped CFI. The point marked with an orange star, in the middle of the gapless region, is studied in Appendix~\ref{app:compareWang} and Sec.~\ref{subsec:examples}.}
    \label{fig:isotropic_CFI_multiE}
\end{figure}

The results of the multi-energy LSA with this isotropy assumption are shown in Fig.~\ref{fig:isotropic_CFI_multiE}. We restrict our attention to the innermost zone where collisions can have a significant impact. The CFI growth rate is shown in the top panel, while the real part of $\Omega'_\mathrm{max}$ is displayed in the bottom panel. There are clearly regions of gapless CFI (dark blue, $\Re(\Omega'_\mathrm{max}) \simeq 0$) and gapped CFI (yellow, $|\Re(\Omega'_\mathrm{max})| \simeq |\Delta_\N - \bDelta_\N| \sim 10^{10} \ \mathrm{s}^{-1}$). The transitions between these regions are associated with larger growth rates (see the darker lines on the top panel), because they correspond to a ``resonancelike" regime~\cite{Liu:2023pjw} where the number densities of $\nu_e$ and $\bar{\nu}_e$ are very close (see top left panel of Fig.~\ref{fig:input_data}). However, we will show in the following that these CFI regions are likely to harbor electron lepton number (ELN) crossings, such that a FFI will overwhelm even a resonancelike CFI. Overall, the CFI, estimated by this approximate isotropy assumption, appears to occur in large regions of the NSM snapshot, especially in the tidal arms, and just above and below the HMNS.

\begin{figure*}[!ht]
    \centering
    \includegraphics[width=\textwidth]{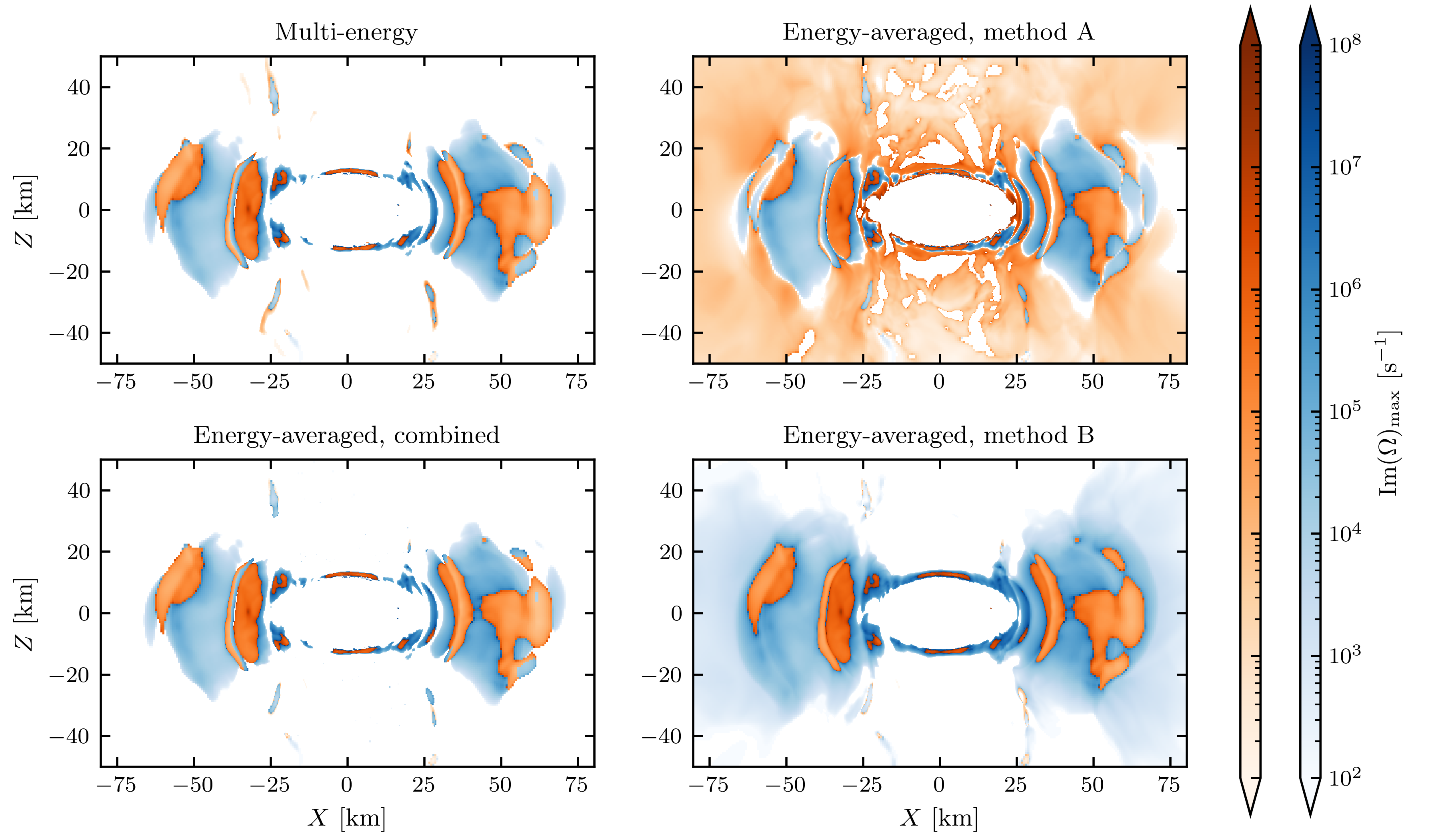}
    \caption{Comparison of various determination of the isotropic and homogeneous CFI growth rate: multi-energy analysis (top left), single-energy analysis with averaged collision rates via method A (top right) and method B (bottom right). Two color maps are used for the gapped (blue) and gapless (orange) modes. In the bottom left panel, the combination of monochromatic results~\eqref{eq:combination_minusA_plusB} shows excellent agreement with the multi-energy results.}
    \label{fig:isotropic_CFI_allmethods}
\end{figure*}

%#######%
\subsubsection{Comparison with energy-averaged approaches}
%#######%

In order to save computation time and avoid constructing the full multi-energy LSA framework we presented in Sec.~\ref{subsec:LSA}, it would be particularly convenient to be able to use the analytic monochromatic formulas describing the CFI (see Appendix~\ref{app:CFI_iso}), as done for instance in~\cite{Froustey:2023skf,Akaho:2023brj,Nagakura:2025hss}. Can the monochromatic results, used with properly energy-averaged quantities, reproduce the multi-energy results?

This question was recently answered in the negative in Ref.~\cite{Wang:2025vbx} for the case of CCSNe, where it was shown that the energy-averaging methods previously used (see Sec.~\ref{subsec:energy_averaging}) largely overestimated the occurrence of CFI. While NSMs have comparable conditions to CCSNe, the differences are such that the same conclusion does not necessarily hold. To determine the answer for NSMs, we compare in Fig.~\ref{fig:isotropic_CFI_allmethods} the different LSA approaches. The top left panel is the multi-energy result, identical to the top panel of Fig.~\ref{fig:isotropic_CFI_multiE} but where we use different color schemes for the gapped (blue) and gapless (orange) regions. We adopt the same color convention for the energy-averaged approaches: method A [see Eq.~\eqref{eq:methodA}] in the top right panel and method B [see Eq.~\eqref{eq:methodB}] in the bottom right panel.

\begin{figure}[!ht]
    \centering
    \includegraphics[width=0.95\columnwidth]{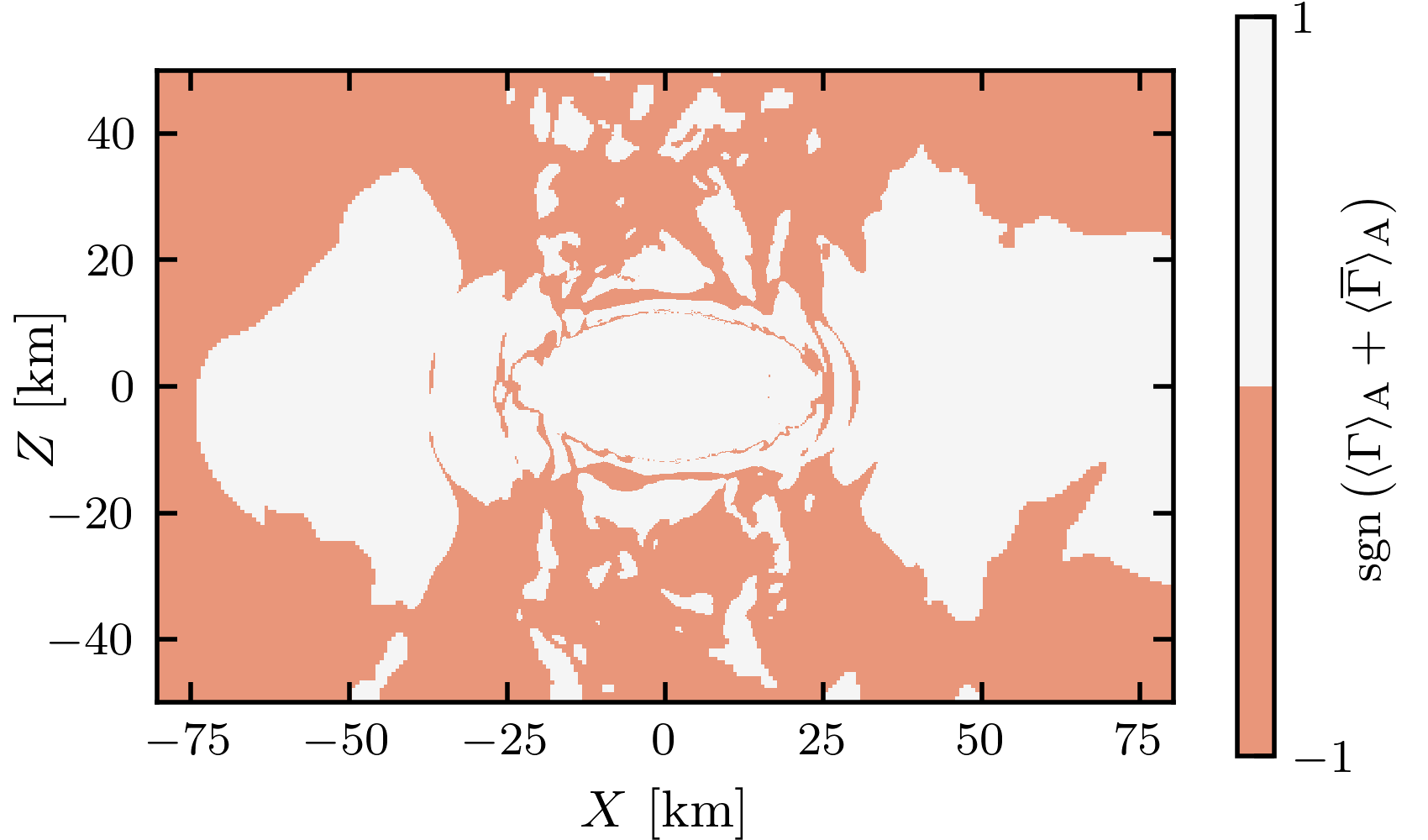}
    \caption{Sign of the sum of the energy-averaged damping rates $\langle \Gamma \rangle_\mathrm{A} + \langle \bGamma \rangle_\mathrm{A}$ for the same slice as Fig.~\ref{fig:isotropic_CFI_allmethods}.}
    \label{fig:sign_gamma_A}
\end{figure}

\paragraph*{Qualitative agreement} 
We have shown in Eq.~\eqref{eq:multi_gapped} that we expect the gapped regions to be correctly described by method A. This is indeed the case (compare the blue regions of the top panels of Fig.~\ref{fig:isotropic_CFI_allmethods}). However, method A predicts very large regions of gapless instability that are not present in the multi-energy analysis. We highlight in particular that the averaging method of Eq.~\eqref{eq:methodA} does not guarantee that the effective damping rates $\langle \Gamma \rangle_\mathrm{A}$ and $\langle \bGamma \rangle_\mathrm{A}$ are positive. We illustrate this in Fig.~\ref{fig:sign_gamma_A}, where we show the sign of $\langle \Gamma \rangle_\mathrm{A} + \langle \bGamma \rangle_\mathrm{A}$ (we note that this expression is dominated by $\langle \Gamma \rangle_\mathrm{A}$). There are very large regions of negative energy-averaged damping rates, which coincide with the fictitious gapless modes visible in the top right panel of Fig.~\ref{fig:isotropic_CFI_allmethods}.

We can understand the origin of the negative $\langle \Gamma \rangle_\mathrm{A}$ regions by noting that the opacities scale with the energy as $\Gamma(p) \simeq \Gamma_0 (p/p_0)^2$, where $p_0$ is an arbitrary reference neutrino energy. With this scaling and using the distributions~\eqref{eq:distrib_gray}, we can rewrite
\begin{equation}
    \langle \Gamma \rangle_\mathrm{A} \propto \frac{\Gamma_0}{p_0^2}\frac{g_{\nu_e} T_{\nu_e}^5 - g_{\nu_x} T_{\nu_x}^5}{\N_{ee} - \N_{xx}} \propto \frac{\Gamma_0}{p_0^2}\frac{\N_{ee} T_{\nu_e}^2 - \N_{xx}T_{\nu_x}^2}{\N_{ee} - \N_{xx}} \, ,
\end{equation}
such that the sign of $\langle \Gamma \rangle_\mathrm{A}$ can be deduced from the bottom left and top right panels of Fig.~\ref{fig:input_data}. A negative sign arises, for $\N_{ee} > \N_{xx}$, when the average energy of $\nu_x$ is sufficiently large to overcome the difference of densities, $(T_{\nu_x}/T_{\nu_e})^2 > \N_{ee}/\N_{xx}$, which leads to fictitious instabilities. This cannot happen with method B, since the weighing function in~\eqref{eq:gray_opacity} is strictly positive.

Method B significantly overestimates the gapped regions, but it provides an excellent estimate of the regions of multi-energy gapless instability. This can be understood with an order-of-magnitude argument. The limit of instability for the gapless B modes corresponds to $\langle \Gamma \rangle_\mathrm{B} \bDelta_\N = \langle \bGamma \rangle_\mathrm{B} \Delta_\N$, see Appendix~\ref{app:gapped_gapless} and especially Eq.~\eqref{eq:plus_mode} for details. We can then make two key approximations: the heavy-lepton flavor neutrinos interact much less with the matter, $\kappa_x \ll \kappa_e, \bar{\kappa}_e$; and the number density of $\nu_x = \bar{\nu}_x$ is small compared to the ones of $\nu_e$ and $\bar{\nu}_e$. Note that this second assumption is violated in CCSNe, which would invalidate the following argument (we expand on this point, looking at a specific example and comparing with Ref.~\cite{Wang:2025vbx}, in Appendix~\ref{app:compareWang}). Furthermore, it is not valid everywhere in the NSM snapshot, but it applies in the regions of gapless instability (compare Figs.~\ref{fig:input_data} and \ref{fig:isotropic_CFI_multiE}). Under the aforementioned assumptions, and using the approximate energy dependence of the opacities $\kappa_e(p) \simeq \kappa_0 (p/p_0)^2$, we have
\begin{equation}
\label{eq:gammaB_approx}
    \langle \Gamma \rangle_\mathrm{B} \simeq \frac{\kappa_0}{2 p_0^2} \frac{\int{p^4 \dd{p} \, f_{\nu_e}(p)}}{\int{p^2 \dd{p} \, f_{\nu_e}(p)}} \propto \frac{\kappa_0}{p_0^2} T_{\nu_e}^2 \, ,
\end{equation}
where we used the gray spectrum~\eqref{eq:distrib_gray}. We also have $\Delta_\N \propto g_{\nu_e} T_{\nu_e}^3$. Doing the same thing for antineutrinos, the instability threshold is
\begin{equation}
\label{eq:limit_stability_plus_B}
    \kappa_0 \, g_{\bar{\nu}_e} T_{\bar{\nu}_e} = \bar{\kappa}_0 \, g_{\nu_e} T_{\nu_e} \, .
\end{equation}
On the other hand, the limit of stability of the multi-energy gapless modes is given by Eq.~(7.9) in~\cite{Fiorillo:2025zio}, which can be written with our notations
\begin{equation}
\label{eq:7.9Fiorillo}
    \int_{0}^{\infty}{p^2 \dd{p} \frac{f_{\nu_e}(p)-f_{\nu_x}(p)}{\Gamma(p)}} = \int_{0}^{\infty}{p^2 \dd{p} \frac{f_{\bar{\nu}_e}(p)-f_{\bar{\nu}_x}(p)}{\bGamma(p)}} \, .
\end{equation}
Under our simplifying assumptions, it reads
\begin{equation}
    \frac{1}{\kappa_0}\int_{0}^{\infty}{p^2 \dd{p} \frac{g_{\nu_e}}{e^{p/T_{\nu_e}}+1}} = \frac{1}{\bar{\kappa}_0}\int_{0}^{\infty}{p^2 \dd{p} \frac{g_{\bar{\nu}_e}}{e^{p/T_{\bar{\nu}_e}}+1}} \, ,
\end{equation}
which corresponds exactly to the condition~\eqref{eq:limit_stability_plus_B}. This explains why the orange regions in the top left and bottom right panels of Fig.~\ref{fig:isotropic_CFI_allmethods} match closely.

The reader might be surprised by the fact that, if there are no $\nu_x$, methods A and B are identical---therefore, why would the previous argument not also apply to the gapless modes of method A? The reason is that “neglecting $\nu_x$” is not equivalent in both methods. In the derivation above, we neglect $\nu_x$ in several places. First, in $\Delta_\N$, which requires $g_{\nu_e} T_{\nu_e}^3 \gg g_{\nu_x} T_{\nu_x}^3$. In the multi-energy instability criterion~\eqref{eq:7.9Fiorillo}, since $\Gamma \propto p^2$ appears in the denominator, we need $g_{\nu_e} T_{\nu_e} \gg g_{\nu_x} T_{\nu_x}$. If we were to use method A, when computing $\langle \Gamma \rangle_\mathrm{A}$ with Eq.~\eqref{eq:methodA}, since $\Gamma$ appears in the numerator, we would need $g_{\nu_e} T_{\nu_e}^5 \gg g_{\nu_x} T_{\nu_x}^5$ before we are able to neglect the $\nu_x$ contribution. However, since $T_{\nu_x} > T_{\nu_e}$, but not by too much (see top left panel of Fig.~\ref{fig:input_data}), it is possible to neglect the $\nu_x$ contribution in Eqs.~\eqref{eq:gammaB_approx}--\eqref{eq:7.9Fiorillo}, but it is less straightforward in the expression of $\langle \Gamma \rangle_\mathrm{A}$, and this is what leads to the negative sign encountered in Fig.~\ref{fig:sign_gamma_A}.

\begin{figure}[!ht]
    \centering
    \includegraphics[width=0.95\columnwidth]{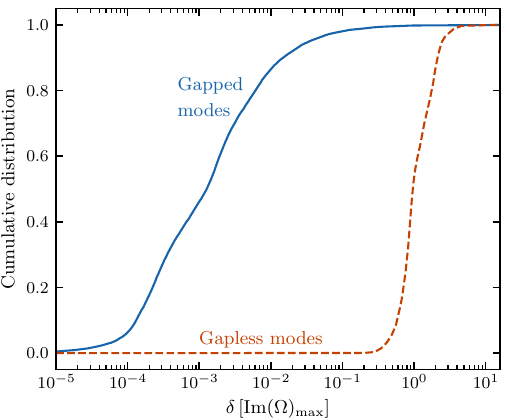}
    \caption{Cumulative distribution function of the relative difference between the CFI growth rates obtained with method A (evaluated on gapped modes) or method B (evaluated on gapless modes), and the multi-energy result.}
    \label{fig:performance_AB}
\end{figure}

\begin{figure}[!ht]
    \centering
    \includegraphics[width=0.95\columnwidth]{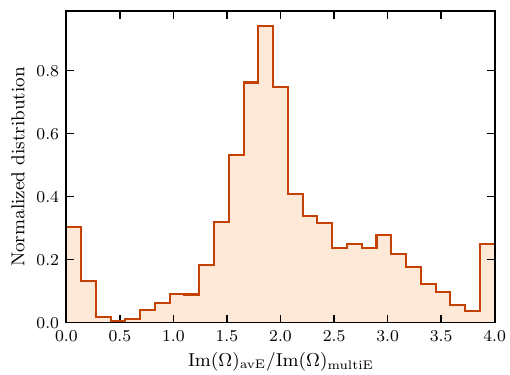}
    \caption{Distribution of the ratio of the growth rates estimated with the energy-averaging method B and the multi-energy results. The growth rate is, on average, typically overestimated by a factor 2.}
    \label{fig:histogram_methodB}
\end{figure}

\paragraph*{Quantitative comparison}
To assess more quantitatively the performance of the energy-averaged approaches, we show in Fig.~\ref{fig:performance_AB} the cumulative distribution of the difference between the monochromatic estimate and the multi-energy result, $\delta[\Im(\Omega)_\mathrm{max}] = |\Im(\Omega)_\text{avE}/\Im(\Omega)_\text{multiE} - 1|$, distinguishing between gapped and gapless regions. For the gapped regions, we compare the multi-energy growth rate with method A, proving that the relative difference is smaller than a percent for $80 \%$ of the grid points, as expected. For the gapless regions we use the method B monochromatic result. The performance is clearly worse than in the gapped case; however, the relative difference is typically of order unity. This is confirmed by looking more specifically at the ratio between the method B prediction and the actual multi-energy gapless CFI growth rate. The histogram of this ratio is shown in Fig.~\ref{fig:histogram_methodB}. Although there is some dispersion, we see that for the conditions encountered in this NSM snapshot, there is typically a factor 2 overestimation of the growth rate by method B. This is very different from what was observed in CCSN environments in~\cite{Wang:2025vbx}, and we attribute it to the hierarchy of number densities $\N_{xx} = \overline{\N}_{xx} \ll \N_{ee}, \overline{\N}_{ee}$ (see Appendix~\ref{app:compareWang}).

Finally, we note that in the monochromatic case away from the resonance regime, gapped and gapless instabilities are mutually exclusive, as can be seen by comparing Eqs.~\eqref{eq:minus_mode} and \eqref{eq:plus_mode}: assuming for instance $\Delta_\N > \bDelta_\N$, $\Im(\Omega'_-) > 0$ if $\Gamma_\N/\bGamma_\N < \bDelta_\N/\Delta_\N < 1$, while $\Im(\Omega'_+) > 0$ if $\Gamma_\N/\bGamma_\N > \Delta_\N/\bDelta_\N > 1$. However, there is no such mutual exclusion in the multi-energy case, and we have found instances where both a gapless and a gapped mode are unstable. This can be identified when simultaneously the gapped mode of method A and the gapless mode of method B are unstable (i.e., a point in Fig.~\ref{fig:isotropic_CFI_allmethods} in blue in the top right panel and in orange in the bottom right panel). In this situation, the multi-energy result shown in the top left panel corresponds to the largest growth rate among the two. Since we know that the gapped method A growth rate matches the multi-energy gapped one, and the gapless method B growth rate must be typically divided by 2, we build the following monochromatic estimate of the multi-energy CFI:
\begin{equation}
    \label{eq:combination_minusA_plusB}
    \sigma_\mathrm{CFI} \simeq \max \left\{\Im(\Omega'_-)_\mathrm{A} \, , \ \frac12 \Im(\Omega'_+)_\mathrm{B} \right\} \, .
\end{equation}
This combination is depicted in the bottom left panel of Fig.~\ref{fig:isotropic_CFI_allmethods}, and shows excellent qualitative and quantitative agreement with the multi-energy results. 

We caution the reader that the regions of instability shown in Fig.~\ref{fig:isotropic_CFI_allmethods} may be slightly misleading, since they are obtained assuming isotropic neutrino distributions. In particular, we will show in Sec.~\ref{sec:results_aniso} that in many of these unstable regions a FFI also takes place which will have a larger growth rate. However, when we restrict our analysis to the regions of the snapshot where there is no angular crossing,
the combination of ``gapped A modes + rescaled gapless B modes" again performs well compared to a full multi-energy analysis.

%-##-%
\subsection{Many-body corrections to the absorption rates}
\label{subsec:manybody}
%-##-%

The absorption opacities that enter into the stability analysis can be altered by many-body corrections that stem from the interaction of the neutrinos with the many-nucleon system.\footnote{We emphasize that these corrections have nothing to do with the possible neutrino quantum many-body correlation effects (see, e.g.~\cite{Patwardhan:2022mxg,Balantekin:2023ayx}), and which we do not consider here.} These corrections can be calculated using linear response theory i.e. by considering the response of the many-nucleon system to the perturbation caused by an incoming neutrino. Interested readers are referred to~\cite{Burrows:1998cg,Burrows:2004vq,Reddy:1997yr,Reddy:1998hb,Roberts:2016mwj,Horowitz:2016gul,Lin:2022lug,Lin:2025glw,Shin:2023sei} for a full explanation of these many-body corrections, and we summarize below their main features.

These corrections are jointly determined by the kinematics of neutrino-nucleon reactions, the thermal distribution, the dispersion relation and the residual interactions of nucleons, which all vary with the nucleon density, local temperature, proton fraction and incoming neutrino energies. At low densities ($\lesssim10^{10}\,\mathrm{g \, cm^{-3}}$), the many-body corrections are not sensitive to the nucleon interactions and are mainly determined by the kinematics.
%Also note that the PS correction introduced by kinematics is different from the one discussed in \cite{horowitz:2001xf}, where the transferred energy is always assumed to be sharply peaked. 
Correctly applying the kinematics automatically includes the phase space corrections, which depend on the incoming neutrino energy and the temperature, and are slightly different for $\nu_e$ and $\bar{\nu}_e$. At medium ($10^{10}\,\mathrm{g \, cm^{-3}} \lesssim \rho \lesssim10^{13}\,\mathrm{g \, cm^{-3}}$) and high ($\rho\gtrsim10^{13}\,\mathrm{g \, cm^{-3}}$) densities, the nucleon-nucleon interactions noticeably alter the correlations of the nucleonic matter and change the neutrino opacity. Here, we apply the Hartree-Fock + random phase approximation (HF+RPA) as in~\cite{Lin:2025glw} to describe the correlation effects. The mean-field (MF) effect described by HF is mainly caused by the difference of neutron and proton potentials, which increases (decreases) the opacity of $\nu_e$ ($\bar{\nu}_e$). On the other hand, the RPA effect significantly modifies the differential neutrino-nucleon cross sections and may result in a resonance peak. Recent studies~\cite{Shin:2023sei,Lin:2022dek} show that in NSM- and CCSN-like conditions, the RPA effect may decrease (increase) the opacity of $\nu_e$ ($\bar{\nu}_e$), countering the MF effect. At medium densities, the MF and the RPA effect are approximately comparable, resulting in a relatively weak net effect. At high densities, the MF effect may dominate the RPA effect at low temperature and low proton fraction, changing the opacities by several orders of magnitude. When the proton fraction is approaching $50 \, \%$, the increase (decrease) of $\kappa_{\mathrm{a},e}$ ($\bar{\kappa}_{\mathrm{a},e}$) due to the MF effect is significantly weakened. At high temperature (e.g., $T\gtrsim 20 \, \mathrm{MeV}$), the many-body correction is much less important than at low temperature. The standard opacities based on Bruenn~\cite{Bruenn:1985en} include a simplified description of many-body corrections which may  overestimate the reduction of $\kappa_{\mathrm{a},e}, \, \bar{\kappa}_{\mathrm{a},e}$ at high temperatures, such that in that regime our modified opacities are larger than the standard ones.

\begin{figure}[!ht]
    \centering
    \includegraphics[width=0.97\columnwidth]{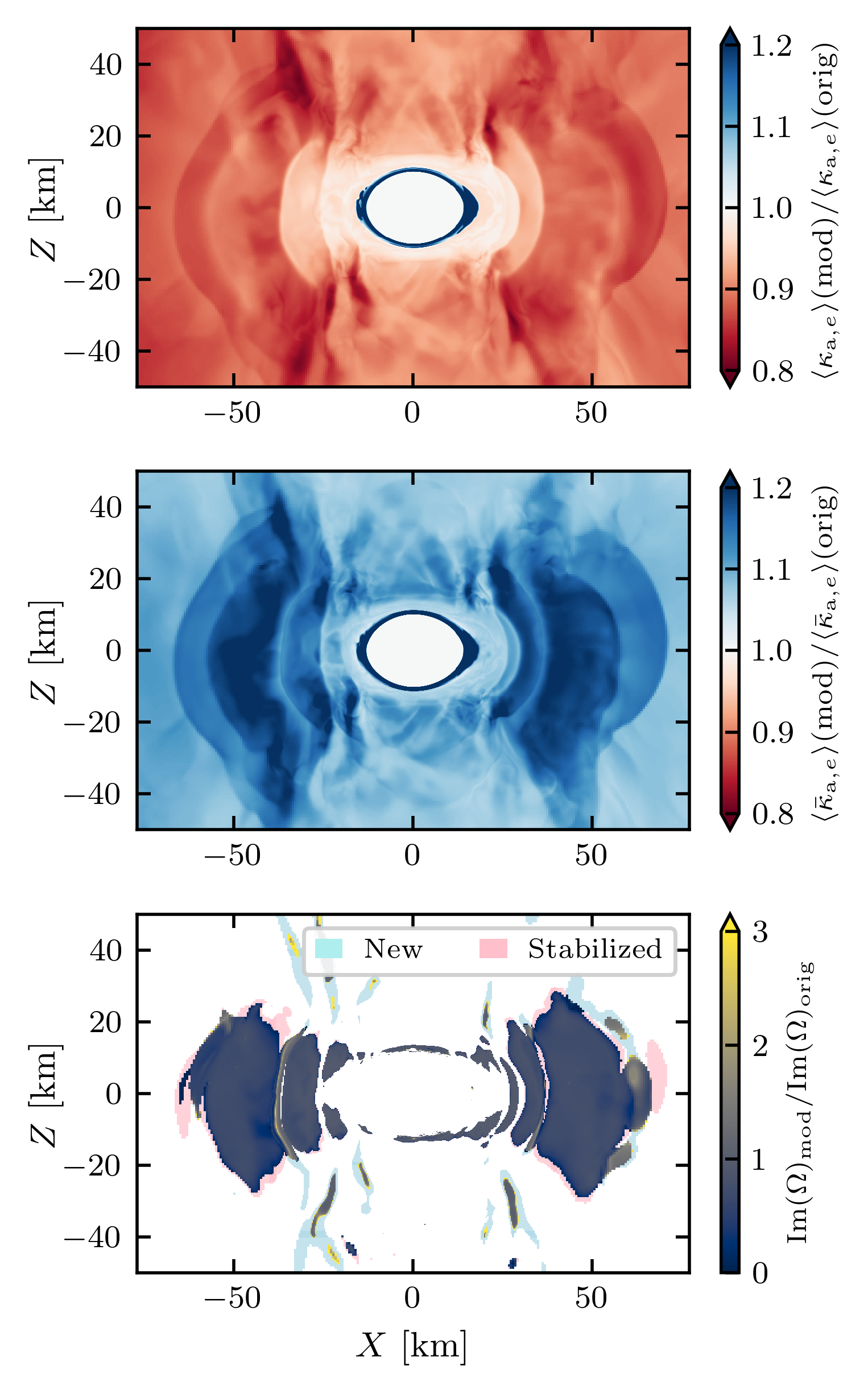}
    \caption{Effect of many-nucleon corrections on the absorption opacities and the homogeneous and isotropic CFI in the 7 ms postmerger snapshot. \emph{Top:} ratio of modified (``mod") and original (``orig") energy-averaged opacities [Eq.~\eqref{eq:gray_opacity}] for $\nu_e$. \emph{Middle:} same for $\bar{\nu}_e$. \emph{Bottom:} ratio of CFI growth rates with modified and original opacities across the snapshot. The original CFI growth rates are shown in the top panel of Fig.~\ref{fig:isotropic_CFI_multiE}. The turquoise color represents new regions of instability originating from the modified opacities, and the pink color represents stabilization of previously unstable regions. Note that we used the dispersion relation approach discussed in Appendix~\ref{app:dispersion_relation} to find the instabilities. It agrees perfectly with the stability matrix results (see also the comparison in Sec.~\ref{subsec:comparison_matrix_disp}).}
    \label{fig:mod_op}
\end{figure}

These effects are at the origin of the changes in opacities depicted in the first two panels of Fig.~\ref{fig:mod_op}. In order to represent in a single quantity the net effect of the many-body corrections, we plot the ratio of the gray opacities as defined in Eq.~\eqref{eq:gray_opacity}. Note that at densities above the saturation density, we do not calculate the many-body correction (white region in the HMNS), since neutrinos are in classical equilibrium, for which there is no instability~\cite{Liu:2024wzd}. For electron neutrinos (top panel), we notice a general trend of decrease in $\langle \kappa_{\mathrm{a},e} \rangle$, except for a sharp increase close to the HMNS---but this happens in a region that is far from collisional instability. We also notice sharp changes in the radial direction, which correspond to sharp changes in the thermodynamic quantities, associated with the tidal arms around the HMNS. For electron antineutrinos, we get a similar pattern but with generally increasing opacities, and a sharp change in the ratio around the same region as for the electron neutrino opacities. Near the remnant and when the density is still small enough that we calculate the correction, the increase in $\langle \bar{\kappa}_{\mathrm{a},e} \rangle$ is consistent with the mechanisms outlined above in the high-temperature regime.

We check how these corrections modify the collisional instability landscape by plotting in the bottom panel of Fig.~\ref{fig:mod_op} the ratio of the CFI growth rate with and without many-body corrections (compare with Fig.~\ref{fig:isotropic_CFI_multiE}). We can see that for the 7 ms snapshot we study, these corrections do not significantly alter the unstable regions. Some very low unstable growth rates, near the fringes of instability, seem to vanish (pink color), and some sparse new unstable regions appear (turquoise color). The ratios of the modified to original growth rates for the rest of the unstable regions are mostly $< 1$, but within order of unity. Since, in the snapshot, neutrino opacities are mostly larger than the antineutrino opacities (see bottom right panel of Fig.~\ref{fig:input_data}), the decrease (increase) in $\langle {\kappa}_{\mathrm{a},e} \rangle$ ($\langle \bar{\kappa}_{\mathrm{a},e} \rangle$), reduces the discrepancy between the collision rates, and thus subdues the instability. These relatively small changes in growth rates are consistent with the variations of opacities that we can see in the CFI regions, which are typically below $\sim 20 \, \%$ (see top and middle panels of Fig.~\ref{fig:mod_op}). Since there are no dramatic changes due to the many-body corrections, for consistency with the collision rates that were used in the NSM simulation from which we extract the data, we will use the original opacities in the rest of this work.
%In later time when the BSM system further cools down, the effect of many-body corrections might be more obvious.

%-##-%
\subsection{Scattering processes and isotropy-breaking modes}
\label{subsec:iso_scatt}
%-##-%

In the previous subsections, we have restricted the collision term to the emission/absorption processes, such that $\Gamma_N = \Gamma_F$. In this limit, the results from monochromatic LSA (see Appendix~\ref{app:CFI_iso}) indicate that the isotropy-breaking modes have smaller growth rates than the isotropy-preserving ones. This was shown numerically in~\cite{Liu:2023pjw}, and we see it directly in the resonancelike regime from Eq.~\eqref{eq:resonancelike}. Away from the resonance, an expansion at second order of Eqs.~\eqref{eq:disp_isopres} and \eqref{eq:disp_isobreak} gives

\begin{equation}
\begin{aligned}
    \mathrm{Im}(\Omega'_\text{pres}) &= -\gamma + \frac{G \alpha}{A} - \frac{G(G^2 - A^2)}{2A^2} \frac{\alpha^3}{A^3} + \mathcal{O}\left(\frac{\alpha^5}{A^5}\right) \, , \\
    \mathrm{Im}(\Omega'_\text{break}) &= -\gamma + \frac{G \alpha}{A} - \frac{9 G(G^2 - A^2)}{2 A^2} \frac{\alpha^3}{A^3} + \mathcal{O}\left(\frac{\alpha^5}{A^5}\right) \, ,
\end{aligned}
\end{equation}
where all the definitions can be found in Eq.~\eqref{eq:notations_Zaizen}; we do not reproduce them here for brevity. We assumed, without loss of generality, $\Delta_\N > \bDelta_\N$ (i.e., $A>0$), and we dropped the indices $\N$ and $\F$ as there is no difference here. Since $|\Delta_\N - \bDelta_\N| \geq \Delta_\N + \bDelta_\N$ (i.e., $|A| \leq G$), we see that the growth rate of the isotropy-breaking modes is smaller. However, the inclusion of isotropic scattering processes, which enter the flux equation but not the number density one [see Eq.~\eqref{eq:collision_N_F}], can lead \emph{a priori} to $- \gamma_\F + (G \alpha_\F)/A > - \gamma_\N + (G \alpha_\N)/A$, if the inevitable increase in $\gamma_\F > \gamma_\N$ is compensated by an increase $\alpha_\F > \alpha_\N$.

\begin{figure}[!ht]
    \centering
    \includegraphics[width=0.97\columnwidth]{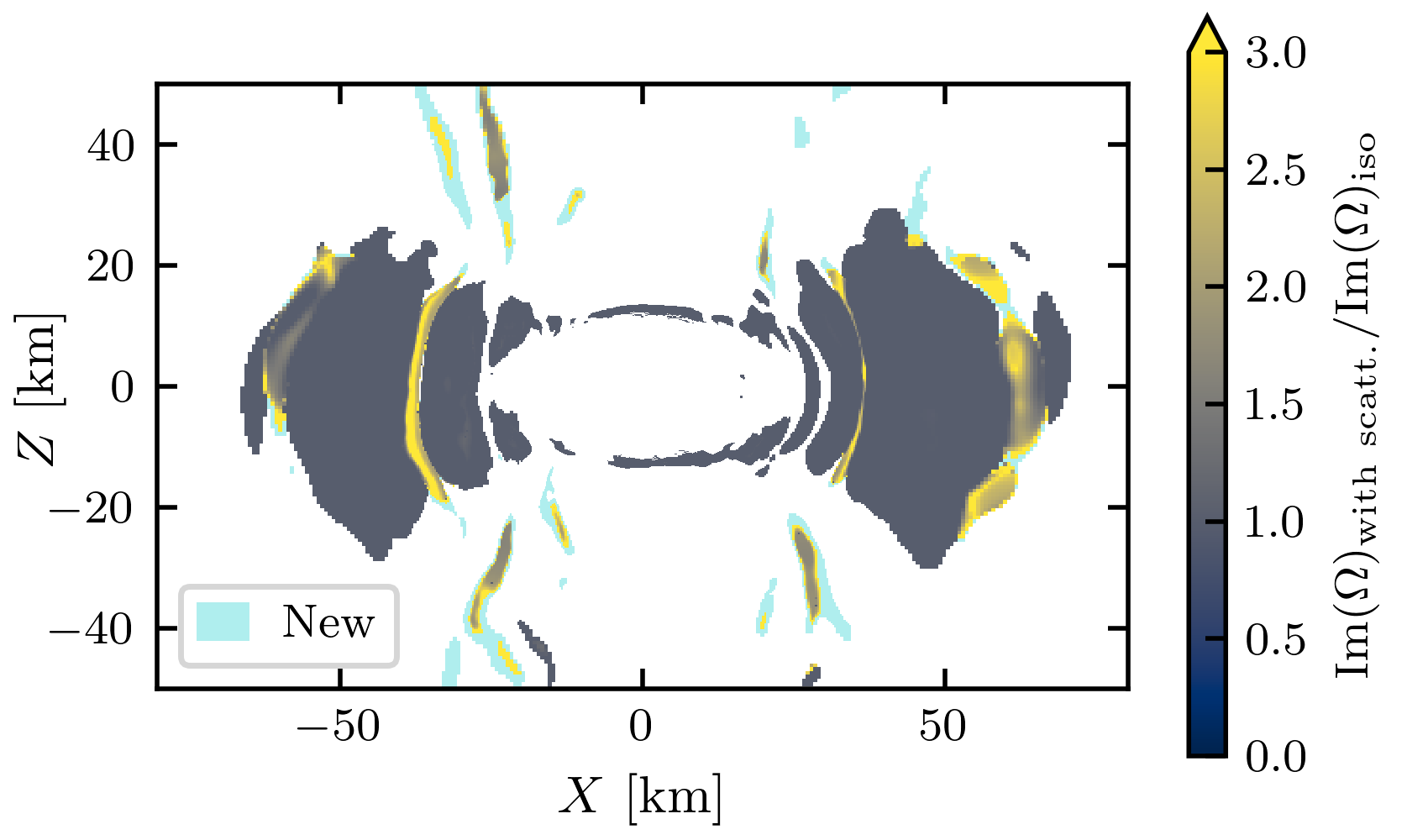}
    \caption{Ratio of the homogeneous CFI growth rate with and without including the scattering opacities in Eq.~\eqref{eq:collision_N_F}. We assume that the flavor on-diagonal distributions are isotropic, such that the scattering processes can intervene via the isotropy-breaking modes~\eqref{eq:disp_isobreak}. The regions in turquoise were not unstable in the absorption-only situation.}
    \label{fig:isoscatt}
\end{figure}

We study this possibility by performing the same multi-energy LSA as in Sec.~\ref{subsec:isotropic_CFI}, but taking into account the scattering opacity contribution to the flux collision term, $\Gamma_F$. Note that the on-diagonal neutrino distributions are still assumed to be isotropic, contrary to what we will do in Sec.~\ref{sec:results_aniso}. We show in Fig.~\ref{fig:isoscatt} the ratio of the CFI growth rate with and without scattering processes, the latter corresponding to the top panel of Fig.~\ref{fig:isotropic_CFI_multiE}. Although most of the instability regions see no significant change in the growth rate, some regions (which corresponded to the smallest growth rates) see an increase in the growth rate by a factor of a few. We also show in turquoise the newly unstable regions, which are typically located around the regions showing a quantitative increase of $\Im(\Omega)_\mathrm{max}$ when one includes the scattering opacities.

The morphology of the changes is similar to what we obtained when considering many-body corrections to the absorption opacities (Fig.~\ref{fig:mod_op}). We attribute this to the fact that the main changes occur in regions which are marginally (un)stable, such that small changes can have a visible impact.

%-##-%
\subsection{Vacuum term and slow modes}
\label{subsec:iso_vac}
%-##-%

In this section, we still assume that the classical neutrino distributions are isotropic and neglect scattering processes, but we include the contribution from the vacuum Hamiltonian.

\begin{figure}[!ht]
    \centering
    \includegraphics[width=0.97\columnwidth]{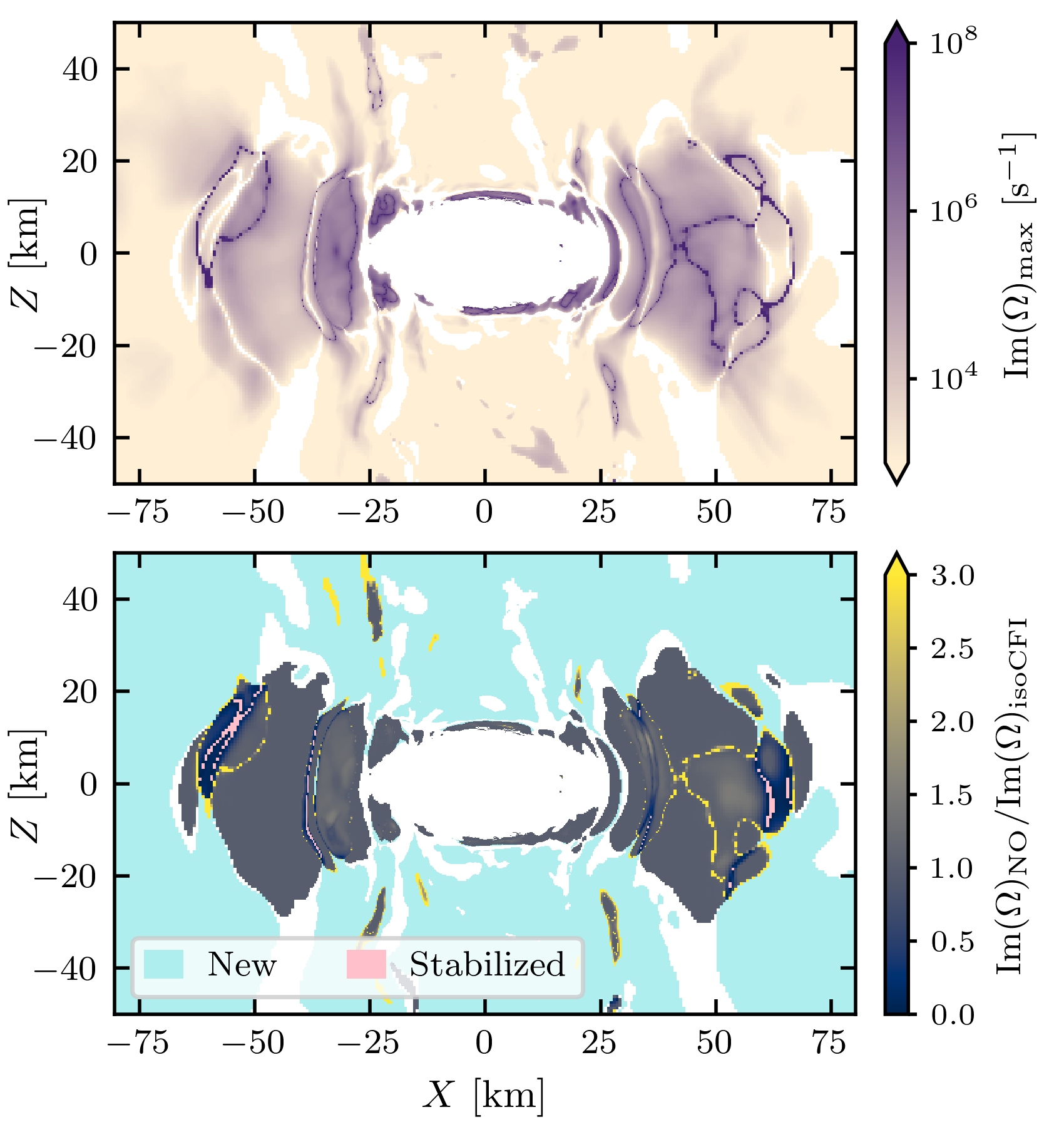}
    \caption{Homogeneous and isotropic instability growth rate in the $Y=0$ slice of the NSM snapshot we study, to be compared with the top panel of Fig.~\ref{fig:isotropic_CFI_multiE}. We assumed here a normal ordering of neutrino masses ($\Delta m^2 = 2.5 \times 10^{-3} \, \mathrm{eV}^2$).}
    \label{fig:isoNO}
\end{figure}

For regions with a gapped CFI, we can generalize the expansion~\eqref{eq:expansion_gapped} to include the vacuum term. This amounts to the replacement $\ii \Sigma_0 \to \ii \Sigma_0 + S_0$, where
\begin{equation}
    S_0 = \sqrt{2} G_F \int_{0}^{\infty}{\frac{p^2 \dd{p}}{2 \pi^2}\left[(f_{\nu_e} - f_{\nu_x}) + (f_{\bar{\nu}_e} - f_{\bar{\nu}_x})\right] \, \omega(p)} \, .
\end{equation}
Therefore, the imaginary part of Eq.~\eqref{eq:multi_gapped} is not modified. However, in the regions with a gapless or no CFI, slow modes could play a role. We show the growth rates obtained from homogeneous and isotropic LSA in Fig.~\ref{fig:isoNO}, top panel. They result from a combination of collisional and slow modes. In the bottom panel, we display the ratio of this growth rate with the default isotropic CFI-only (top panel of Fig.~\ref{fig:isotropic_CFI_multiE}). As in Fig.~\ref{fig:mod_op}, we show in turquoise the newly unstable regions, and in pink the regions which were unstable without the vacuum term but are stable with it. As expected from the argument above, the regions with the largest differences in growth rates correspond to the regions of gapless instability or of stability in Fig.~\ref{fig:isotropic_CFI_multiE}. The results of Fig.~\ref{fig:isoNO} assume a normal ordering of neutrino masses---the results for the inverted ordering are very similar and we do not show them for brevity. The newly unstable regions are mostly associated with very small growth rates ($< 10^4 \, \mathrm{s}^{-1}$), such that we do not expect a large impact of these slow modes. In particular, local studies of the slow instability in those regions would not be justified, since large-scale advection occurs on comparable or smaller timescales. Nevertheless, as we will show in the following, including the anisotropies of the neutrino field can lead to a significant enhancement of these growth rates (see Sec.~\ref{subsec:fullpicture}).

%%%%%%%%%%%%%%%%%%%
\section{Results for anisotropic neutrino distributions}
\label{sec:results_aniso}
%%%%%%%%%%%%%%%%%%%

In this section and for the remainder of this paper, we no longer assume that neutrino distributions are isotropic, and we include the actual nonzero fluxes $\vec{F}_{\alpha \alpha} \neq \vec{0}$. This has two consequences. First, angular crossings between the neutrino and antineutrino distributions can occur, which are connected to fast instabilities~\cite{Morinaga:2021vmc,Dasgupta:2021gfs,Fiorillo:2024bzm}. Second, the separation between isotropy-preserving and isotropy-breaking modes---which can be done when the stability matrix has a block structure; see Eq.~\eqref{eq:S_hom_iso}---is not possible anymore. Furthermore, we also now consider different collision rates for number and flux, $\Gamma_N$ and $\Gamma_F$.

\begin{figure*}[!ht]
    \centering
    \includegraphics[width=0.9\textwidth]{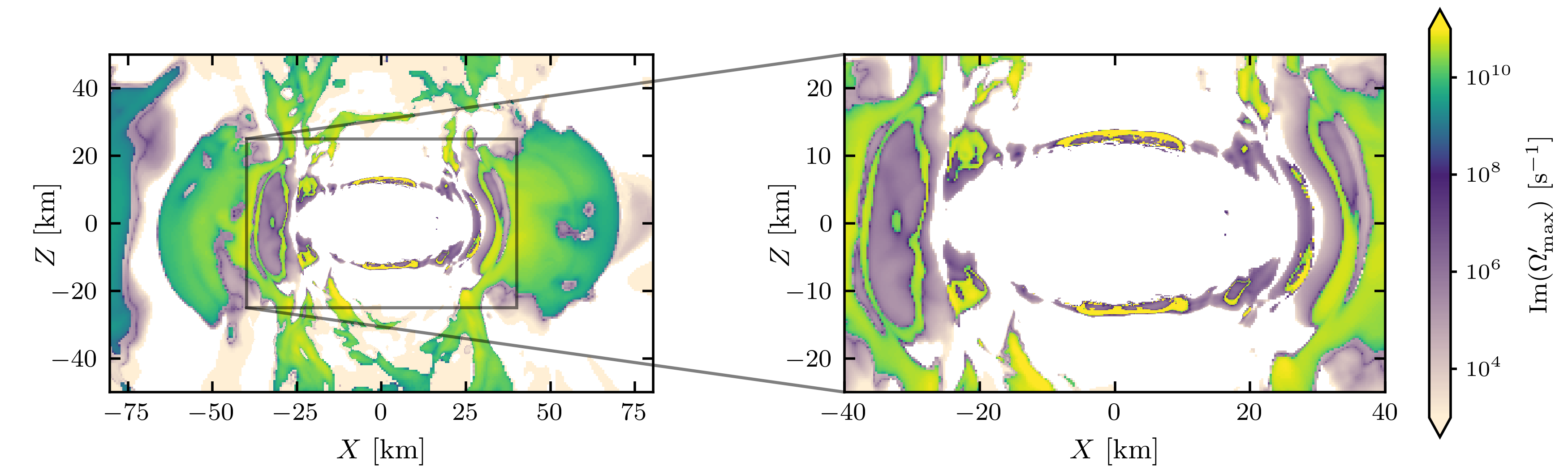}  
    \caption{Result of the multi-energy moment LSA at $\vec{k}' = \vec{0}$. We use the same color bar as the top panel of Fig.~\ref{fig:isotropic_CFI_multiE}, extended for large values of $\Im(\Omega')_\mathrm{max}$ to describe the FFI.}
    \label{fig:anisotropic_momentLSA_multiE}
\end{figure*}

%---------------------%
\subsection{Results at $k'=0$, no vacuum term}
\label{subsec:aniso_zeromode}
%---------------------%

We show in Fig.~\ref{fig:anisotropic_momentLSA_multiE} the results of our multi-energy moment LSA restricted to $\vec{k}' = \vec{0}$, without including the vacuum term. This allows us to focus on the occurrence of fast and collisional modes. Comparing the left panel with the top panel of Fig.~\ref{fig:isotropic_CFI_multiE} shows that most of the isotropic CFI regions at large distances ($|X| \geq 40 \, \mathrm{km}$) are actually regions of FFI. Even though it is not surprising that the isotropy assumption was particularly incorrect in such locations, it is worth noting that angular crossings occur in those regions. In particular, the ``resonancelike" regions, for which $\N_{ee} \simeq \overline{\N}_{ee}$, are particularly prone to harbor crossings, since for similar densities any difference in flux factor or flux direction will lead to a crossing. We thus conclude that the resonancelike CFI is a negligible phenomenon in the NSM snapshots we have studied. 

Although our goal here is not to discuss in depth the origin of the angular crossings that lead to FFI, we note that our regions of FFI are consistent with other works. First, the instability patterns (small regions above and below the HMNS, extended regions in the tidal arms) are similar to the ones found in a 5 ms postmerger snapshot of another binary NSM simulation in~\cite{Froustey:2023skf} (see also Appendix~\ref{app:other_snapshots}). Another comparison can be made with the recent detailed study~\cite{Nagakura:2025hss}, which used multiangle Boltzmann neutrino transport on top of fixed fluid backgrounds obtained from a $1.35 M_\odot - 1.35 M_\odot$ NSM simulation leading to a long-lived ($> 1.3 \, \mathrm{s}$) HMNS. While in the simulation we study the HMNS collapses to a black hole at $8.5 \, \mathrm{ms}$ post-merger, some of the same physical mechanisms occur. For instance, the thin layers near the surface of the HMNS correspond to near-isotropic configurations with similar number densities of $\nu_e$ and $\bar{\nu}_e$, such that a small difference in flux between these two species leads to a crossing (see Sec.~III.B.1. in~\cite{Nagakura:2025hss}). Reference~\cite{Nagakura:2025hss} points out other mechanisms, mostly based on the larger emission of $\bar{\nu}_e$ than $\nu_e$ at the surface of the HMNS, and the presence of an optically thick accretion disk with $\mu_\nu^\mathrm{eq} > 0$. However, some of the processes leading to crossings are dependent on having sufficient angular resolution, and may be smeared out by using a two-moment scheme assuming maximum entropy angular distributions.

\begin{figure*}[!ht]
    \centering
    \includegraphics[width=0.9\textwidth]{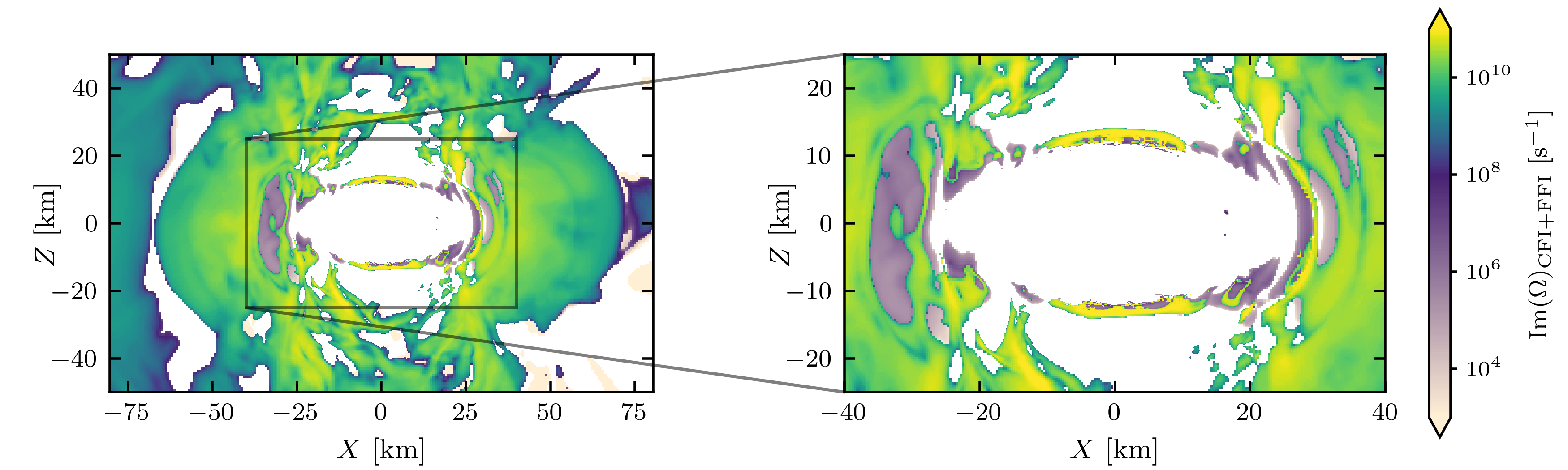}
    \caption{Combination of the multi-energy moment LSA at $\vec{k}' = \vec{0}$ with the FFI estimate~\eqref{eq:approx_FFI}, showing all the regions where an analysis not limited at $\vec{k}' = \vec{0}$ would lead to a growth rate largely surpassing an underlying CFI.}
    \label{fig:anisotropic_CFIFFI_multiE}
\end{figure*}

\begin{figure*}[!ht]
    \centering
    \includegraphics[width=0.9\textwidth]{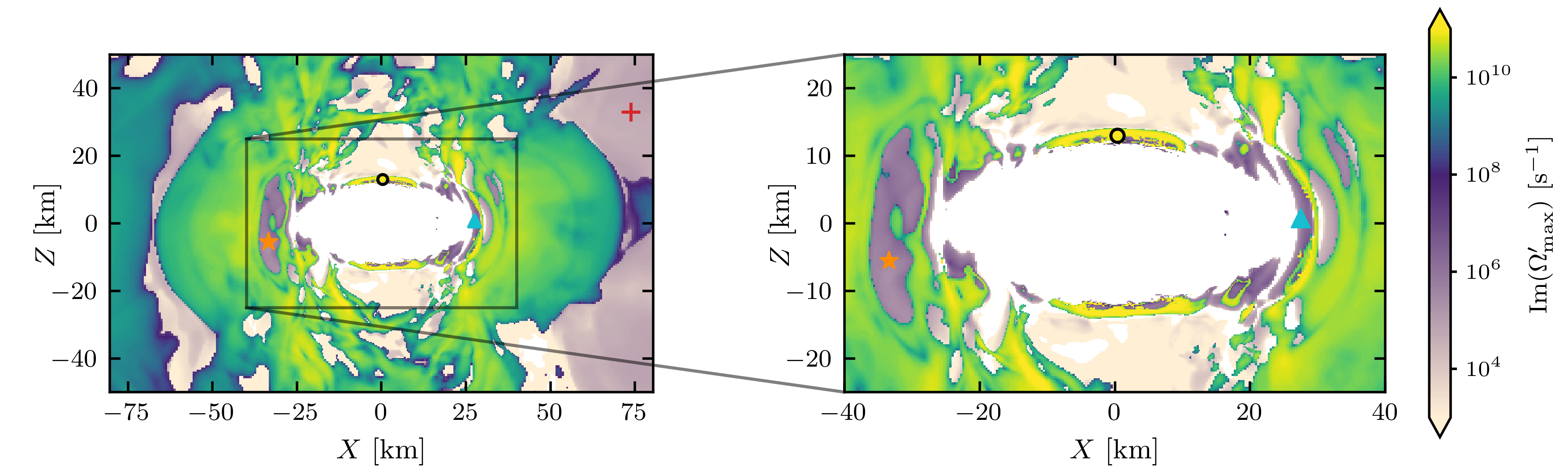}
    \caption{Combination of the multi-energy moment LSA at $\vec{k}' = \vec{0}$, including the vacuum term (for normal ordering), with the FFI estimate~\eqref{eq:approx_FFI}. We identify a few points that are specifically discussed in Sec.~\ref{subsec:examples}.}
    \label{fig:allFI_multiE}
\end{figure*}

Since the CFI is subdominant in the regions far from the HMNS, we focus upon the central region; see the right panel of Fig.~\ref{fig:anisotropic_momentLSA_multiE}. Contrary to what was observed in CCSNe~\cite{Wang:2025vbx}, we find that the CFI is still largely present, extending the regions of instability. The growth rates are significantly smaller, but they could still lead to a buildup of flavor coherence throughout the equatorial region. The light-crossing time of a $1 \, \mathrm{km}$ region is $\tau \simeq 1/(3 \times 10^5) \, \mathrm{s}$, which shows that instabilities with growth rates below a few times $10^5 \, \mathrm{s}^{-1}$ can only have a cumulative, nonlocal effect (contrary to the FFI), which remains to be studied. We also caution that our LSA assumes constant and homogeneous flavor on-diagonal densities, such that our results only give us information on the local properties of the instabilities. Although this is beyond the scope of this work, we refer to~\cite{Shalgar:2023aca} for a discussion of a possible way to perform a nonlocal LSA which takes into account the large-scale inhomogeneities of the classical neutrino field.

%---------------------%
\subsection{Fast, collisional and slow modes: Toward the full picture}
\label{subsec:fullpicture}
%---------------------%

A limitation of our analysis, restricted to the zero mode $\vec{k}' = \vec{0}$, is to underestimate the occurrence of the FFI (for an illustration, see Fig.~8 in~\cite{Froustey:2023skf}). For the CFI, in the isotropic case Ref.~\cite{Liu:2023pjw} has found that the $k=0$ mode had the largest growth rate. Whether this remains true with anisotropic distributions in the absence of a crossing is less clear, but we will take our results as a conservative estimate of the CFI. Since we want to describe the additional regions of instability which appear with collisions, underestimating CFI regions is not as much of a problem as underestimating FFI regions.

In order to make sure that every region with a crossing is predicted to be unstable, we can use the empirical formula that connects the growth rate to the depth of the crossing~\cite{Morinaga:2019wsv}:
\begin{equation}
\label{eq:approx_FFI}
    \sigma_\mathrm{FFI} \simeq \sqrt{2} G_F \sqrt{I_+ I_-} \, ,
\end{equation}
with $I_+$ and $I_-$ given by
\begin{equation}
\begin{aligned}
    I_+ &\equiv \int_{f_{\nu_e} - f_{\bar{\nu}_e} > 0}{\frac{\dd^3 \vec{p}}{(2 \pi)^3} \left[f_{\nu_e}(\vec{p}) - f_{\bar{\nu}_e}(\vec{p})\right]} \, , \\
    I_- &\equiv \int_{f_{\nu_e} - f_{\bar{\nu}_e} < 0}{\frac{\dd^3 \vec{p}}{(2 \pi)^3} \left[f_{\bar{\nu}_e}(\vec{p}) - f_{\nu_e}(\vec{p}) \right]} \, .
\end{aligned}
\end{equation}
Note we have assumed $f_{\nu_x} = f_{\bar{\nu}_x}$. In these expressions, we have used the ansatz~\eqref{eq:distrib} for the neutrino spectra, where the angular part corresponds to maximum entropy distributions~\eqref{eq:distrib_ME}. Although the formula~\eqref{eq:approx_FFI} may not be quantitatively exact, it ensures a FFI growth rate of order $\sqrt{2} G_F \N \sim 10^{10} \, \mathrm{s}^{-1}$---much larger than the CFI growth rates---for all regions where a crossing should be present (assuming maximum entropy angular distributions). We thus superimpose this estimate on the results at $\vec{k}' = \vec{0}$, see Fig.~\ref{fig:anisotropic_CFIFFI_multiE}. In the left panel, we see that now almost all CFI regions at large distances from the HMNS are overwhelmed by a FFI, but the enlarged version (right panel) still shows an extension of the instability regions in thin layers near the surface of the HMNS and, more significantly, in the tidal arms. We note that a detailed comparison of the FFI regions between Figs.~\ref{fig:anisotropic_momentLSA_multiE} and \ref{fig:anisotropic_CFIFFI_multiE} shows that the expression \eqref{eq:approx_FFI} can underestimate the FFI growth rate since the $\vec{k}'=\vec{0}$ mode does not necessarily have the largest growth rate. 

Beyond the quantitative uncertainty of Eq.~\eqref{eq:approx_FFI}, the Minerbo closure is an imperfect proxy for neutrino angular distributions. It was notably shown in Ref.~\cite{Richers:2022dqa} that, in NSM environments, instability regions vary depending on the reconstruction method used compared to benchmark multiangle Monte-Carlo calculations. Likewise, the use of multiangle Boltzmann transport in the CCSN simulations of Akaho \emph{et al.}~\cite{Akaho:2026kff} allowed them to highlight the errors incurred with a Minerbo reconstruction for the detection of angular crossings. Despite these limitations, our use of moment-based NSM simulation data necessitates a closure-driven reconstruction. We note that, in the same situation but for CCSN simulations, Wang and Burrows~\cite{Wang:2025ihh} compared the few closure relations available which provide a full angular distribution. They found very similar CCSN dynamics when different angular reconstruction methods were used in a FFI subgrid model.

Finally, we combine in Fig.~\ref{fig:allFI_multiE} the moment LSA results at $\vec{k}' = \vec{0}$, \emph{including the vacuum term}, and the FFI estimate~\eqref{eq:approx_FFI}. The additional regions of instabilities between Figs.~\ref{fig:anisotropic_CFIFFI_multiE} and \ref{fig:allFI_multiE} are driven by the vacuum term and can be classified as ``slow modes." They mainly occur in the polar regions, above the thin slabs of FFI/CFI above and below the HMNS, and at large distances (see for instance the region $X \geq 50 \, \mathrm{km}$). Importantly, we see that the growth rates in these distant regions are much larger than in the isotropic case (see Fig.~\ref{fig:isoNO}). Whether there are modes with even larger growth rates at $\vec{k}' \neq \vec{0}$ in such locations remains to be seen and is a question for future work. The instability map shown in Fig.~\ref{fig:allFI_multiE} is our most complete estimate of the occurrence of the instabilities in a NSM postmerger snapshot, combining fast, collisional, and slow modes.

%---------------------%
\subsection{Summary: Focus on a few example locations}
\label{subsec:examples}
%---------------------%

\begin{figure}[!ht]
    \centering
    \includegraphics[width=0.97\columnwidth]{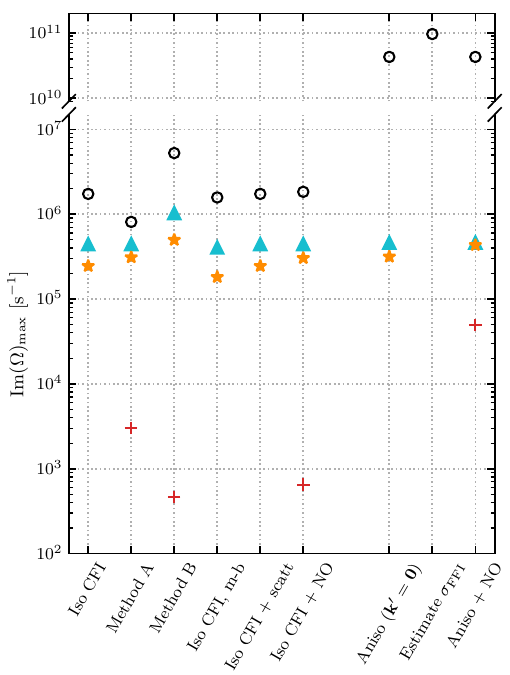}
    \caption{Instability growth rate for some points identified in Fig.~\ref{fig:allFI_multiE}, under the different assumptions considered in this work (see text). Black circles represent a point that is unstable to the FFI, the red cross is unstable to slow modes, the blue triangle is unstable to the gapped CFI, and the orange star is unstable to the gapless CFI. The absence of a point means that the result was negative, hence showing no instability for that particular assumption.}
    \label{fig:summary_fewpoints}
\end{figure}

In order to summarize the several growth rate determinations we have obtained in this work, and how different regions of a NSM simulation are subject to different flavor instabilities, we focus here on four locations identified in Fig.~\ref{fig:allFI_multiE}. The instability growth rate for these four points is shown in Fig.~\ref{fig:summary_fewpoints}, where the marker matches the one used in Fig.~\ref{fig:allFI_multiE} and each column corresponds to a different LSA assumption. From left to right, first assuming isotropic neutrino background distributions, ``Iso CFI" is the multi-energy CFI, compared with the energy-averaged ``Method A"~\eqref{eq:methodA} and ``Method B"~\eqref{eq:methodB} (see Sec.~\ref{subsec:isotropic_CFI}), and with various multi-energy extensions: with many-body corrections to the absorption opacities (``Iso CFI, m-b," Sec.~\ref{subsec:manybody}), including scattering opacities (``Iso CFI + scatt," Sec.~\ref{subsec:iso_scatt}), or with the vacuum term (``Iso CFI + NO," Sec.~\ref{subsec:iso_vac}). Then, we show the results obtained when we take into account the nonzero neutrino fluxes: ``Aniso ($\vec{k}' = \vec{0}$)" when the vacuum term is not included (Sec.~\ref{subsec:aniso_zeromode}), ``Estimate $\sigma_\mathrm{FFI}$" for the estimate~\eqref{eq:approx_FFI}, and ``Aniso + NO" for the result at $\vec{k}'=\vec{0}$ including the vacuum term (Sec.~\ref{subsec:fullpicture}).

The blue triangle and orange star points, located near the HMNS and close to the equatorial plane, show relatively little variation between the predicted growth rates. The blue triangle shows a gapped CFI, such that method A accurately predicts the isotropic CFI growth rate. The many-body corrections or scattering opacities do not make a significant change, just like the vacuum term (which is expected for a gapped mode, as discussed in Sec.~\ref{subsec:iso_vac}). Including the anisotropies makes very little change, which is consistent given the small flux factors for this point (the largest flux factor is $\tilde{f}_x = 0.035$, and the smallest is $\tilde{f}_x = 0.004$). The dark orange point (which is also studied in Appendix~\ref{app:compareWang} in comparison with a CCSN example) displays a gapless CFI, such that method A does not predict exactly the isotropic CFI growth rate, and as we showed in Sec.~\ref{subsec:isotropic_CFI}, method B typically overestimates the growth rate by a factor 2. The many-body corrections reduce the growth rate by a factor of a few, as expected from Fig.~\ref{fig:mod_op}, bottom panel. Including the vacuum term slightly increases the growth rate, an effect that is enhanced in the anisotropic case (here the flux factors range from $\tilde{f}_e = 0.011$ to $\tilde{f}_x = 0.047$).

The black circle point, located in the thin layer of instability above the HMNS, displays a gapless CFI when we neglect the anisotropies, but the configuration is actually fast-unstable. Although Eq.~\eqref{eq:approx_FFI} may seem to overestimate the FFI growth rate, one must remember that the value for $\vec{k}'=\vec{0}$ is a lower bound for the instability growth rate. As expected for a fast instability, the vacuum term has a negligible effect.

The final configuration we look at is located much further away from the HMNS; see the red cross in the left panel of Fig.~\ref{fig:allFI_multiE}. Focusing on the isotropic CFI, the reader may be surprised by the false predictions of methods A and B, when we showed in Sec.~\ref{subsec:isotropic_CFI} that we could accurately estimate the multi-energy growth rate with those methods. However, looking at Fig.~\ref{fig:isotropic_CFI_allmethods} one can see that the unstable energy-averaged modes shown in Fig.~\ref{fig:summary_fewpoints} are a gapless mode for method A and a gapped one for method B. But our accurate estimate~\eqref{eq:combination_minusA_plusB} combines gapped A modes and gapless B modes, which would give here, as expected, no instability. The inclusion of the vacuum term indicates a possible slow mode, but given the large distance of that point from the HMNS, neglecting the anisotropies is not justified ($\nu_e$ and $\bar{\nu}_e$ distributions have flux factors $\tilde{f} \simeq 0.7$). Indeed, we see that even though this point is not unstable to the FFI, the slow mode is enhanced by the anisotropies by about 2 orders of magnitude. This is representative of what happens in the region $X \geq 50 \, \mathrm{km}$ on the left panel of Fig.~\ref{fig:allFI_multiE}. We leave for future work the detailed study of the mechanisms that lead to this anisotropy-driven growth rate enhancement, and their possible interplay with collisions.

%%%%%%%%%%%%%%%%%%%
\section{Conclusion}
\label{sec:conclusion}
%%%%%%%%%%%%%%%%%%%

Linear stability analysis allows one to determine the locations and timescales of neutrino flavor instabilities, which correspond to an exponential growth of flavor coherence when one includes flavor mixing effects in neutrino transport. In this work, we have extended the angular moment-based LSA framework developed in~\cite{Froustey:2023skf} to describe a multi-energy system and included the vacuum and collision terms, which are, respectively, associated with slow and collisional instabilities. In order to avoid the inherent quantum closure problems associated with a moment method~\cite{Myers:2021hnp,Froustey:2024sgz,Kneller:2024buy,Grohs:2025ajr}, we restricted the LSA to the ``zero mode," which is a wavevector coinciding with the homogeneous mode for isotropic background neutrino distributions and for which a classical closure relation is sufficient. We have applied our LSA to a 7 ms postmerger snapshot from the ``M1-NuLib" simulation presented in Ref.~\cite{Foucart:2024npn}.

Because of their analytical simplicity, most estimates of the CFI in dense astrophysical environments have been made with the restrictive assumptions of monochromaticity and isotropy. However, spurred by the recent findings in~\cite{Wang:2025vbx} that the single-energy approximation dramatically overestimates the occurrence of CFI in CCSN environments, we compare multi- and single-energy isotropic estimates in Sec.~\ref{sec:results_isotropic_CFI}. Using the classification from~\cite{Fiorillo:2025zio}, we find that NSM environments can harbor both gapped (modes where $|\Re(\Omega')| \gg \Gamma$) and gapless (modes where $|\Re(\Omega')| \simeq 0$) instabilities, while only the latter occur in CCSNe. Moreover, the energy-averaged methods introduced in the literature work surprisingly well in the NSM environment, provided that one uses different averaging methods in the gapped and gapless cases. Our simplified formula for CFIs in the isotropic case is given in Eq.~\eqref{eq:combination_minusA_plusB} and shows excellent qualitative and quantitative agreement with the multi-energy results (see Fig.~\ref{fig:isotropic_CFI_allmethods}). Still assuming isotropic classical neutrino distributions, we have studied the changes in the CFI due to nucleon many-body corrections to the absorption opacities (Sec.~\ref{subsec:manybody}), although they are subdominant due to the relatively small densities involved in CFI-unstable regions. Our angular moment framework allows one to naturally include both absorption and elastic scattering processes, which act differently on the number and flux densities, although this effect is minute for isotropic distributions (Sec.~\ref{subsec:iso_scatt}). The inclusion of the QKE vacuum terms reveals the presence of slow modes at large distances from the hypermassive neutron star, although with growth rates typically corresponding to $\sim \mathrm{ms}$ and beyond timescales (Sec.~\ref{subsec:iso_vac}).

Since we find many unstable modes at large distances from the HMNS, where neglecting the neutrino fluxes is not justified, we present in Sec.~\ref{sec:results_aniso} a faithful assessment of flavor instabilities for the ``zero,'' $\vec{k}' = \vec{0}$, mode, where the anisotropies of the neutrino field are taken into account. For the first time, we include not only absorption but also elastic scattering processes in the equations of motion in a multi-energy setting. To compensate for the underestimation of FFIs (which have the largest growth rates) due to the restricted wavenumber, we also include a growth rate estimate based on the depth of the FFI-associated angular crossing [Eq.~\eqref{eq:approx_FFI}], which ensures that only slow and collisional modes can be missed. We find that the NSM snapshot we studied is mostly dominated by the FFI (see e.g.,~\cite{Nagakura:2025hss} for a discussion of the mechanisms leading to FFI in a similar system but with a long-lived HMNS), which in particular occurs in the regions where a resonancelike CFI could have taken place. Nevertheless, gapless and gapped CFIs still occur and extend the range of fast-flavor unstable regions. Finally, the slow modes at large distances appear to be enhanced by the anisotropies (compare Figs.~\ref{fig:isoNO} and \ref{fig:allFI_multiE}, and see the red cross points in Fig.~\ref{fig:summary_fewpoints}). Overall, Fig.~\ref{fig:allFI_multiE} (and the bottom panels of Figs.~\ref{fig:summary_M13ms} and \ref{fig:summary_OldM1} for other NSM snapshots) represents our most complete estimate of neutrino flavor instabilities in classical moment-based NSM simulation snapshots. In addition to flavor instabilities, other flavor conversion mechanisms can occur in NSM environments, such as MSW, turbulence-driven or matter-neutrino resonances~\cite{Volpe:2023met}. For instance, the latter arise from the cancellation of the matter and self-interaction mean-field potentials~\cite{Malkus:2012ts,Malkus:2014iqa,Wu:2015fga,Frensel:2016fge,Tian:2017xbr,Shalgar:2017pzd,Vlasenko:2018irq,Padilla-Gay:2024wyo} and could have significant effects on nucleosynthesis yields~\cite{Malkus:2012ts,Malkus:2015mda}. However, probing this mechanism requires tracking the propagation of neutrinos through the spatially varying potentials, and the interplay between these phenomena and flavor instabilities remains to be investigated.

As emphasized above, our moment LSA is restricted to the zero mode $\vec{k}' = \vec{0}$, which avoids quantum closure issues and possible spurious modes at the expense of an incomplete assessment of the instabilities. Determining adequate closures for the generic $\vec{k}$ case, for instance based on multiangle LSA, is left for future work. It is also worth remembering that an assessment of flavor instabilities on a classically computed simulation snapshot is not self-consistent, since instabilities would have modified the neutrino distributions in the simulation had they been included. An intermediate step toward an improved diagnosis of flavor instabilities in these environments could be to apply the framework presented here to simulations that already incorporate flavor conversion subgrid physics, such as~\cite{Wang:2025nii,Qiu:2025kgy,Lund:2025jjo,Wang:2025ihh,Qiu:2025ybw}. Finally, we note that although highly useful, a LSA does not provide information on the final outcome of flavor instabilities. In particular for CFIs, although the post-instability asymptotic state has only been studied recently in simplified setups~\cite{Zaizen:2025ptx,Froustey:2025nbi}, the smaller growth rates render almost unavoidable larger-scale simulations where advection and the time evolution of the flavor-diagonal components of the density matrices are taken into account. This is a challenging problem, for which moment approaches, with their inherently lower computational cost, present significant advantages.

%%%%%%%%%%%%%%%%%%%%%%
\begin{acknowledgments}
We thank Evan Grohs, Damiano Fiorillo, and Tianshu Wang for valuable discussions, with special thanks to Tianshu Wang for generously providing the data used in Appendix~\ref{app:compareWang}. J.F. acknowledges support from the Severo Ochoa Excellence Grant CEX2023-001292-S funded by MICIU/AEI/10.13039/501100011033 and from the Network for Neutrinos, Nuclear Astrophysics and Symmetries (N3AS), through the National Science Foundation Physics Frontier Center Award No.~PHY-2020275. J.P.K. and G.C.M. are supported by the United States Department of Energy, Office of Science, Office of Nuclear Physics under Awardx No.~DE-FG02-02ER41216 and No.~DE-SC0024388 (ENAF). G.C.M. also acknowledges support from DOE Contracts No.~LA22-ML-DE-FOA-2440 and No.~DE-AC52-107NA27344.
F.F. is supported by the Department of Energy, Office of Science, Office of Nuclear Physics, under Contract No.~DE-SC0020435 and by NASA through Grant No.~80NSSC22K0719.
\end{acknowledgments}

\appendix

%%%%%%%%%%%%%%%%%%%%%%%%%%%%%%%%%%%%%%%%%%%%%%%%%%%%%
\appsection{Monochromatic and isotropic distributions}
\label{app:CFI_iso}
%%%%%%%%%%%%%%%%%%%%%%%%%%%%%%%%%%%%%%%%%%%%%%%%%%%%%

Previous studies used simplifying assumptions to get analytic dispersion relations describing the CFI. We show here how these results emerge in our moment formalism, providing a clear physical picture of the nature of the unstable modes. Specifically, the neutrino distributions constituting the background of the LSA are assumed to be \emph{isotropic}, see Eq.~\eqref{eq:isotropic_background}. Furthermore, we assume that the system can be described with a \emph{monochromatic} energy spectrum (single-energy approximation). If we call $n_0$ the index of the only occupied energy bin, we have $\N_{ee} = \Delta p_{n_0} N_{ee}^{(n_0)}$, and likewise for all energy-integrated quantities.

%%%%
\subsection{Eigenvalues of the stability matrix}
\label{subsec:isobreak_pres}
%%%%

With these assumptions, the now $8 \times 8$ stability matrix is split into four independent blocks. If we order the vector of perturbation amplitudes as $ \left(\mathcal{A}_{ex},\bar{\mathcal{A}}_{xe},\mathcal{B}_{ex}^\x,\overline{\mathcal{B}}_{xe}^\x,\mathcal{B}_{ex}^\y,\overline{\mathcal{B}}_{xe}^\y,\mathcal{B}_{ex}^\z,\overline{\mathcal{B}}_{xe}^\z\right)^T$, the matrix $\mathsf{S}$ has a block structure given by
\begin{equation}
\label{eq:S_hom_iso}
    \mathsf{S} = \left(\begin{array}{c|c|c|c}
 \mathsf{S}_\text{pres} & 0 & 0 & 0  \\ \hline
0  & \mathsf{S}_\text{break} & 0 & 0 \\ \hline 0 & 0 & \mathsf{S}_\text{break} & 0 \\ \hline  0 & 0 & 0 & \mathsf{S}_\text{break}
 \end{array} \right) \, .
\end{equation}
The subscripts “pres” and “break” refer to the “isotropy-preserving” and “isotropy-breaking” modes, respectively, following the nomenclature of~\cite{Liu:2023pjw,Liu:2023vtz,Akaho:2023brj}. For consistency with those works, we restrict to CFI and take $\omega=0$ in this section unless otherwise specified.

\paragraph*{$(\mathcal{A}_{ex},\bar{\mathcal{A}}_{xe})$ subspace ---} The $2 \times 2$ submatrix corresponding to the number density perturbations is
\begin{equation}
\label{eq:S_isopres}
    \mathsf{S}_\text{pres} = \begin{pmatrix} - \Delta_\N - \ii \, \Gamma_\N & \Delta_\N \\ - \bDelta_\N &  \bDelta_\N - \ii \, \bGamma_\N \end{pmatrix} \, ,
\end{equation}
where $\Gamma_\N$ stands for $\Gamma_N^{(n_0)}$. The eigenvalues of this $2\times2$ matrix can be readily found. For comparison with previous studies, we introduce here the notations~\cite{Liu:2023pjw,Liu:2023vtz,Akaho:2023brj}\footnote{Note that we have the correspondence $\mathfrak{g} = \Delta_\N$ and $\bar{\mathfrak{g}} = \bDelta_\N$.}
\begin{equation}
\label{eq:notations_Zaizen}
\begin{aligned}
    G &\equiv \frac{\Delta_\N + \bDelta_\N}{2} \, ,  & \gamma_\N &\equiv \frac{\Gamma_\N + \bGamma_\N}{2} \, , \\
    A &\equiv \frac{\Delta_\N - \bDelta_\N}{2} \, ,  & \alpha_\N &\equiv \frac{\Gamma_\N - \bGamma_\N}{2} \, .
\end{aligned}
\end{equation}
The eigenvalues of $\mathsf{S}_\text{pres}$ are thus
\begin{equation}
\label{eq:disp_isopres}
    \Omega'_\text{pres} = - A - \ii \, \gamma_N \pm \sqrt{A^2 + 2 \, \ii \, G \, \alpha_\N - \alpha_\N^2} \, ,
\end{equation}
in complete agreement with~\cite{Liu:2023pjw,Liu:2023vtz,Akaho:2023brj}. These modes are called “isotropy-preserving” as they are pure \emph{number density} perturbations, such that the neutrino field remains isotropic.

\paragraph*{$(\mathcal{B}_{ex}^i,\overline{\mathcal{B}}_{xe}^i)$ subspace ---} The other submatrices give a solution with multiplicity $3$. With $\mathsf{S}_\text{break}$ given by
\begin{equation}
\label{eq:S_isobreak}
    \mathsf{S}_\text{break} = \begin{pmatrix} \dfrac{\Delta_\N}{3} - \ii \, \Gamma_\F & - \dfrac{\Delta_\N}{3} \\ \dfrac{\bDelta_\N}{3} & - \dfrac{\bDelta_\N}{3} - \ii \, \bGamma_\F \end{pmatrix} \, ,
\end{equation}
the eigenvalues are
\begin{equation}
\label{eq:disp_isobreak}
    \Omega'_\text{break} = \frac{A}{3} - \ii \, \gamma_\F \pm \sqrt{\left(\frac{A}{3}\right)^2 - \frac{2}{3} \, \ii \, G \, \alpha_\F - \alpha_\F^2} \, .
\end{equation}
The quantities $\gamma_\F$ and $\alpha_\F$ are defined in the same fashion as~\eqref{eq:notations_Zaizen}. These modes correspond to nonzero $(\bm{\mathcal{B}}_{ex},\overline{\bm{\mathcal{B}}}_{xe})$, which therefore “break” the isotropy.

\paragraph*{“(Non)resonancelike” CFI ---} NSM environment conditions are usually such that $A^2 \gg \abs{G \alpha_{\N,\F}}$. In that limit, we have
\begin{equation}
    \max \Im(\Omega'_\text{pres}) \simeq \max \Im(\Omega'_\text{break}) \simeq - \gamma + \abs{\frac{G\alpha}{A}} \, ,
\end{equation}
where we dropped the subscripts $\N,\F$ on $\gamma$ and $\alpha$ for brevity. However, this must not hide the fact that the scattering rates that enter $\Gamma_\F$ but not $\Gamma_\N$ will lead to \emph{a priori} different growth rates (see Sec.~\ref{subsec:iso_scatt}). This is the “nonresonance” case~\cite{Akaho:2023brj}. The growth rate can be much larger in the “resonancelike” case, where electron neutrino and antineutrino densities are very close, such that $A \simeq 0$. In this limit $A^2 \ll \abs{G \alpha_{N,F}}$, we have:
\begin{equation}
\label{eq:resonancelike}
\begin{aligned}
    \max \Im(\Omega'_\text{pres}) &\simeq - \gamma_N + \sqrt{\abs{G \alpha_N}} \, , \\
    \max \Im(\Omega'_\text{break}) &\simeq - \gamma_F + \sqrt{\frac{\abs{G \alpha_F}}{3}} \, .    
\end{aligned}
\end{equation}

When including the vacuum Hamiltonian, the stability matrix in the homogeneous and isotropic case keeps the structure~\eqref{eq:S_hom_iso}, where each $2 \times 2$ submatrix is modified by $\mathsf{S}_\text{pres, break} \to \mathsf{S}_{\text{pres,break}} + \mathrm{diag}(-\omega^{(n_0)},\omega^{(n_0)})$. Therefore, the previous dispersion relations remain valid by trading the previous $\alpha$ with $\alpha - \ii \, \omega^{(n_0)}$.

%%%%
\subsection{Gapped/minus and gapless/plus modes}
\label{app:gapped_gapless}
%%%%

As can be seen in Eqs.~\eqref{eq:disp_isopres} and \eqref{eq:disp_isobreak}, each branch of the instability admits two modes, which have been dubbed in the literature ``plus" and ``minus" modes. Although it may be tempting to associate these modes to the signs $+$ and $-$ appearing in those equations, it would be erroneous since the identification of plus and minus modes depends on the sign of the various quantities (see\footnote{We caution the reader that the expressions in \cite{Froustey:2025nbi} are for the physical frequency $\Omega$, not the shifted quantity $\Omega'$, which differs by its real part.} for instance Appendix A in~\cite{Froustey:2025nbi}). A more physically grounded nomenclature of these modes was used in Ref.~\cite{Fiorillo:2025zio}, introducing the terms ``gapless" and ``gapped" modes.

For clarity, we will focus on the isotropy-preserving branch of the instability. In the nonresonance limit $A^2 \gg |G \alpha_\N|$ (with also $\alpha_\N^2 \ll |G \alpha_\N|$), the two eigenvalues \eqref{eq:disp_isopres} read
\begin{equation}
    \Omega'_\text{pres} \simeq (- A \pm |A|) + \ii \left(- \gamma_N \pm \frac{G \alpha_\N}{|A|}\right) \, .
\end{equation}
Regardless of the sign of $A$, we get the so-called ``minus mode" such that
\begin{equation}
\label{eq:minus_mode}
    \Omega'_- \simeq - \left(\Delta_\N - \bDelta_\N\right) + \ii \frac{\bGamma_\N \bDelta_\N - \Gamma_\N \Delta_\N}{\Delta_\N - \bDelta_\N} \, .
\end{equation}
The main feature of this mode is its very large real part $\Re(\Omega'_-)$. Following the nomenclature of~\cite{Fiorillo:2025zio}, this is a \emph{gapped} mode: indeed, as $\Gamma, \bGamma \to 0$, $\Omega'_-$ does not go to zero.

The second (``plus") mode has a vanishing real part,
\begin{equation}
\label{eq:plus_mode}
    \Omega'_+ \simeq \ii \frac{\Gamma_\N \bDelta_\N - \bGamma_\N \Delta_\N}{\Delta_\N - \bDelta_\N} \, ,
\end{equation}
and since $\Omega'_+ \to 0$ for $\Gamma, \bGamma \to 0$, it is a \emph{gapless} mode. Note that the same classification holds if one includes the vacuum term, as the notion of gapless and gapped depends upon the limit $\Gamma, \bGamma \to 0$ and $\Delta m^2 \to 0$~\cite{Fiorillo:2025zio}.

Another qualitative difference between these two modes lies in the associated eigenvectors: for the minus mode, $(\mathcal{A}_{ex}, \bar{\mathcal{A}}_{xe}) \propto (\Delta_\N, \bDelta_\N)$ while for the plus mode, $(\mathcal{A}_{ex}, \bar{\mathcal{A}}_{xe}) \propto (1, 1)$~\cite{Froustey:2025nbi}. This corresponds to two very different precession-like trajectories of the polarization vectors describing the neutrino state~\cite{Fiorillo:2023ajs,Johns:2025mlm}.

In the main text, we generally adopt the gapped/gapless nomenclature, especially since it extends to the multi-energy analysis.

%%%%%%%%%%%%%%%%%%%%%%%%%%%%%%%%%%%%%%%
\appsection{Dispersion relation approach}
\label{app:dispersion_relation}
%%%%%%%%%%%%%%%%%%%%%%%%%%%%%%%%%%%%%%%

The stability matrix method presented in Sec.~\ref{subsec:LSA} consists of finding, with a numerical diagonalization technique, the eigenvalues and eigenvectors of the matrix $\mathsf{S}$ introduced in Eq.~\eqref{eq:stability_matrix}. An alternative but equivalent method, more suited to obtain analytical results, is called the “dispersion relation” approach~\cite{Izaguirre:2016gsx}. It consists of recasting the linearized QKEs into an integral equation involving the frequency $\Omega'$ (which was called a ``consistency condition" in \cite{Banerjee:2011fj}). This can be done for any wavevector $\vec{k}'$, such that the solution of the integral equation provides the “dispersion relation” $\Omega'(\vec{k}')$. Since any numerical solution of the integral equation requires a discretization in energy bins, the results should be equivalent to the stability matrix approach (see for instance \cite{Wang:2025vbx} where the dispersion relation was used for visualization purposes, but the stability matrix was used to obtain the numerical results). We discuss this equivalence in this appendix, restricting the geometry of the problem for clarity.

\subsection{Axially symmetric distributions}

To simplify, we restrict to axisymmetric distributions along the $\z$ direction. As a consequence, the only nonzero components of the first moments are $N$, $F^{\z}$, $P^{\z \z}$ and\footnote{The definition~\eqref{eq:moment_p_P} imposes the geometric relation $P^{ij} \delta_{ij} = N$.} $P^{\x \x} = P^{\y \y} = (N-P^{\z \z})/2$. Summing the equations~\eqref{eq:linearized_QKEs} over energy bins $n$, we get the relation

\begin{widetext}
\begin{equation}
\label{eq:dispersion_relation_matrix}
    \begin{pmatrix} 1 + I_{0,N} & - I_{1,N} & 0 & 0 \\ I_{1, F} & 1 - I_{2,F} & 0 & 0 \\ 0 & 0 & 1 - \frac12(I_{0,F} - I_{2,F}) & 0 \\ 0 & 0 & 0 & 1 - \frac12 (I_{0,F} - I_{2,F}) \end{pmatrix} \begin{pmatrix} \mathcal{A}_{ex} - \bar{\mathcal{A}}_{xe} \\ \mathcal{B}_{ex}^\z - \overline{\mathcal{B}}_{xe}^\z \\ \mathcal{B}_{ex}^\x - \overline{\mathcal{B}}_{xe}^\x \\ \mathcal{B}_{ex}^\y - \overline{\mathcal{B}}_{xe}^\y \end{pmatrix}
    = 0 \, ,
\end{equation}
\end{widetext}
where the calligraphic quantities represent, like before, the energy-integrated quantities such as $\mathcal{A}_{ex} \equiv \sum_{n} \Delta p_n A_{ex}^{(n)}$, and we introduced

\begin{equation}
\label{eq:discrete_integrals}
    \begin{aligned}
        I_{0,M} &= \sum_{n} \left[\frac{\Delta_N^{(n)} \, \Delta p_n}{\Omega' + \omega^{(n)} + \ii \, \Gamma_M^{(n)}} - \frac{\bDelta_N^{(n)} \, \Delta p_n}{\Omega' - \omega^{(n)} + \ii \, \bGamma_M^{(n)}}\right] \, , \\
        I_{1,M} &= \sum_{n} \left[\frac{\Delta_F^{\z,(n)} \, \Delta p_n}{\Omega' + \omega^{(n)} + \ii \, \Gamma_M^{(n)}} - \frac{\bDelta_F^{\z,(n)} \, \Delta p_n}{\Omega' - \omega^{(n)} + \ii \, \bGamma_M^{(n)}}\right] \, , \\
        I_{2,M} &= \sum_{n} \left[\frac{\Delta_P^{\z\z,(n)} \, \Delta p_n}{\Omega' + \omega^{(n)} + \ii \, \Gamma_M^{(n)}} - \frac{\bDelta_P^{\z\z,(n)} \, \Delta p_n}{\Omega' - \omega^{(n)} + \ii \, \bGamma_M^{(n)}}\right] \, ,
    \end{aligned}
\end{equation}
where $M \in \{N,F\}$. These equations correspond to the discretized versions of Eqs.~(2.16) and~(2.17) in~\cite{Fiorillo:2025zio}, with some sign differences because we focus on the $ex$ component of the QKEs, when the dynamical quantity in \cite{Fiorillo:2025zio}, $\psi_\vec{p}$, corresponds to the $xe$ component.\footnote{Specifically, the connection between our conventions and the conventions of \cite{Fiorillo:2025zio} are $\Omega' \leftrightarrow [- \omega^*]_\text{\cite{Fiorillo:2025zio}}$ (note that they have the same imaginary part), and $I_\ell \leftrightarrow [- \mu I_\ell^*(k=0)]_\text{\cite{Fiorillo:2025zio}}$.} Also, we allow here for different number ($\Gamma_N$) and flux ($\Gamma_F$) collision rates, and we restrict to the zero mode $k' = 0$.

The problem~\eqref{eq:dispersion_relation_matrix} has nonzero solutions for “longitudinal modes” satisfying the integral equation
\begin{equation}
    \label{eq:disp_longitudinal}
    (1+ I_{0,N})(1-I_{2,F}) + I_{1,N} I_{1,F} = 0 \, , 
\end{equation}
and “axial-breaking modes” (which produce a flux transverse to the $\z$ direction), satisfying
\begin{equation}
    \label{eq:disp_axialbreak}
    (I_{0,F} - I_{2,F}) - 2 = 0 \, .
\end{equation}

\subsection{Isotropic neutrino distributions}

Let us now assume, as is usually done in the literature on CFIs in astrophysical environments, that the neutrino distributions are isotropic, and let us also neglect elastic scattering processes. As a consequence, $\Gamma_N = \Gamma_F \equiv \Gamma$, $I_{\ell, N} = I_{\ell, F} \equiv I_\ell$, $\Delta_F = 0$ and $\Delta_P^{ij} = (\Delta_N/3) \delta^{ij}$.

The longitudinal modes~\eqref{eq:disp_longitudinal} have two branches. First, the “isotropy-preserving” modes\footnote{Indeed, inserting this solution in~\eqref{eq:dispersion_relation_matrix} shows that $\mathcal{B}_{ex}^j = 0$, such that the flavor wave has no net flux and preserves the isotropy.} verifying $I_0 = - 1$. Then, the “isotropy-breaking” modes\footnote{For these modes, although the flavor on-diagonal distributions are isotropic, the flavor wave develops a nonzero flux $\bm{\mathcal{B}}_{ex} \neq \vec{0}$.} which satisfy $I_2 = 1$, that is, $I_0 = 3$. Now that there is no preferred direction, the axial-breaking modes~\eqref{eq:disp_axialbreak} are also solutions of $I_0 = 3$. Written in continuous form and discarding the vacuum term, these equations read
\begin{multline}
\label{eq:dispersion_CFI}
    \sqrt{2} G_F \int_{0}^{\infty}{\frac{p^2 \dd{p}}{2 \pi^2}\left[\frac{f_{\nu_e}(p) - f_{\nu_x}(p)}{\Omega' + \ii \, \Gamma(p)} -\frac{f_{\bar{\nu}_e}(p) - f_{\bar{\nu}_x}(p)}{\Omega' + \ii \, \bGamma(p)}\right]} \\ = \begin{cases} -1 \\ 3 \end{cases} \, ,
\end{multline}
which is the standard dispersion relation used in the CFI literature (see, e.g.,~\cite{Lin:2022dek,Liu:2023pjw,Wang:2025vbx}). In the monochromatic limit $f_{\nu_\alpha}(p) \propto \delta(p-p_0)$, we can explicitly solve Eq.~\eqref{eq:dispersion_CFI}, which leads to the expressions \eqref{eq:disp_isopres} (for the $-1$ branch) and \eqref{eq:disp_isobreak} (for the $3$ branch).

\subsection{Comparison of the two methods}
\label{subsec:comparison_matrix_disp}

For a general polychromatic spectrum, it is possible to solve numerically Eq.~\eqref{eq:dispersion_CFI} for $\Omega'$ or, more specifically, the discretized version \eqref{eq:discrete_integrals}--\eqref{eq:disp_axialbreak}. For gapped modes, each term of the sum is of order 1, since $\Re(\Omega') \simeq \sum_n (\bDelta_N^{(n)} - \Delta_N^{(n)}) \Delta p_n$ [see Eq.~\eqref{eq:disp_isopres}, valid for a gapped mode]. This is easy to handle numerically. However, for gapless modes, the denominator of each term is typically of order $\Gamma$ (since the real part of $\Omega'$ is very small compared to the imaginary part, itself of order $\sim \Gamma$). This means that each term in the sum is roughly of order $\sqrt{2} G_F \N / \Gamma \sim 10^4$ and above. Numerically, dealing with a sum of very large terms that must be precisely combined to give $-1$ or $3$ is more difficult. As such, we have found that in a few cases, it was important to provide a good seed to the root solver to avoid missing a gapless instability when using Eq.~\eqref{eq:dispersion_CFI}. Apart from this precaution, finding a root of the integral equation or using the stability matrix method are both reliable methods.

Numerically solving Eq.~\eqref{eq:dispersion_CFI} using the routine \texttt{scipy.optimize.root}~\cite{2020SciPy-NMeth} for a single point takes $\sim$ 1 ms of CPU time (including both the $-1$ and $3$ branches). However, since at each point Eq.~\eqref{eq:dispersion_CFI} can have multiple solutions for $\Omega'$, we can use multiple initial (seed) guesses to capture all the solutions (see also the comment above for possible difficulties with gapless modes). We can use an \emph{ad hoc} set of 10 purely imaginary seed guesses $\ii \left[10^0, 10^1, \dots,10^9\right] \mathrm{s}^{-1}$, at equal logarithmic intervals, that covers the typical range of any growth rate that might occur. This multiple seed guesses method scales up the computation time by one order of magnitude. The diagonalization of the stability matrix (with the routine \texttt{scipy.linalg.eig}~\cite{2020SciPy-NMeth}) for a single point takes $\sim 0.2 \, \mathrm{ms}$ if we restrict to the isotropic case [that is, for the multi-energy version of~\eqref{eq:S_isopres}], and $\sim 8 \, \mathrm{ms}$ for the full stability matrix. The stability matrix method is thus very competitive with the dispersion relation approach, since it includes nonzero fluxes, does not require any guesses and is as reliable for gapped and gapless modes. All computation times discussed here were for 16 energy groups, that is the energy binning of the \texttt{NuLib} table introduced in Sec.~\ref{subsec:data_NSM}.

\begin{figure*}[!ht]
    \centering
    \includegraphics[width=0.95\textwidth]{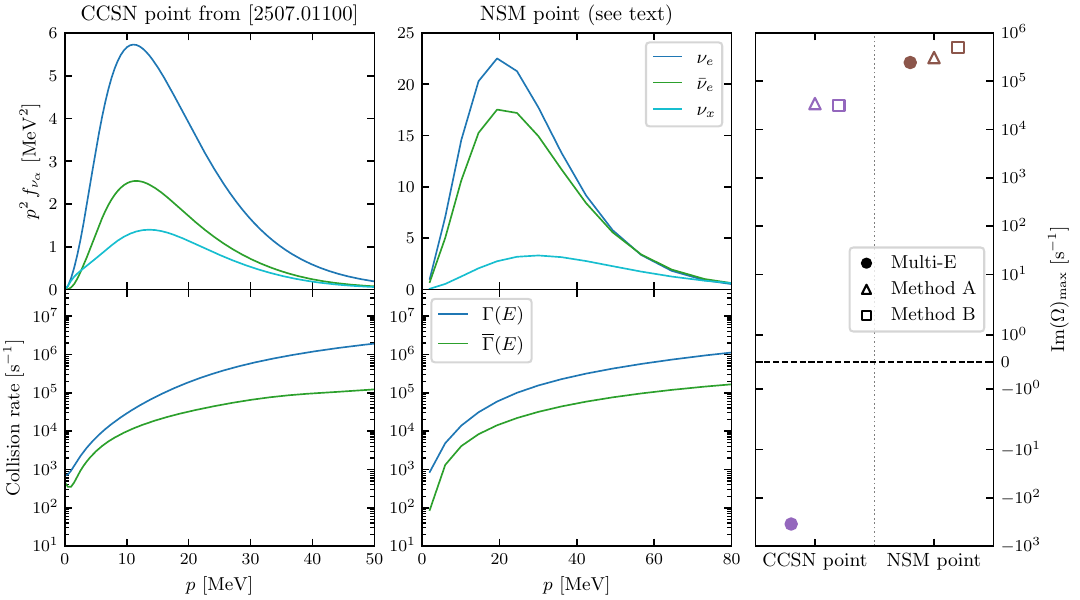}
    \caption{Comparison of typical CFI gapless modes in CCSN and NSM environments. The left panels show the energy spectra and collision rates for the CCSN example, which corresponds to Fig.~1 in \cite{Wang:2025vbx}. The middle panels show the same quantities for the point identified with an orange star in Fig.~\ref{fig:isotropic_CFI_multiE}. In the right panel, we show the result of multi-energy and energy-averaged LSA for these two configurations.}
    \label{fig:comparison_CCSN_NSM}
\end{figure*}

%%%%%%%%%%%%%%%%%%%%%%%%%%%%%%%%%%%%%%%
\appsection{CFI gapless modes in NSM and CCSN environments}
\label{app:compareWang}
%%%%%%%%%%%%%%%%%%%%%%%%%%%%%%%%%%%%%%%

In this appendix, we discuss the origin of the different properties of collisional gapless modes between NSM environments (see Sec.~\ref{sec:results_isotropic_CFI}) and in CCSNe. For the latter, we refer to the work by Wang \emph{et al.}~\cite{Wang:2025vbx}, which found that the energy-averaged approaches overestimated the CFI growth rate by several orders of magnitude, while we find instead an overestimation by a factor $\sim 2$ (see Fig.~\ref{fig:histogram_methodB}).

We can use the monochromatic results (Appendix~\ref{app:CFI_iso}) to get a sense of the kind of unstable modes that can appear. In CCSNe, as pointed out in~\cite{Fiorillo:2025zio}, there are more $\nu_e$ than $\bar{\nu}_e$, so $\Delta_\N > \bDelta_\N$, and the electronic neutrinos interact more, so $\Gamma_\N > \bGamma_\N$. Therefore, we do not expect gapped modes to be unstable, as this would require, per Eq.~\eqref{eq:minus_mode}, $\Gamma_\N / \bGamma_\N < \bDelta_\N / \Delta_\N < 1$. However, a gapless mode can be unstable, if one has $\Gamma_\N / \bGamma_\N > \Delta_\N / \bDelta_\N$.

In the NSM simulation we have studied, we also have typically $\Gamma_\N > \bGamma_\N$ (see bottom right panel of Fig.~\ref{fig:input_data}), but there are regions with more $\bar{\nu}_e$ than $\nu_e$, which allows for gapped modes. There are also regions of gapless instability, where the difference with the multi-energy results is not nearly as dramatic as in~\cite{Wang:2025vbx}.

To highlight the difference between those two astrophysical environments, we focus on two simulation points. The first one is studied in Fig. 1 of Ref.~\cite{Wang:2025vbx} and corresponds to the configuration at a radius of 37 km, at 120 ms post-bounce for a $18 M_\odot$ 1D CCSN simulation. The second point is shown with an orange star in Fig.~\ref{fig:isotropic_CFI_multiE}, it has coordinates $(X,Y,Z)\simeq (-33 \, \mathrm{km}, 0 \, \mathrm{km}, -5 \, \mathrm{km})$. It is chosen because it is in the middle of a gapless region that remains a CFI region even when including anisotropies (see Fig.~\ref{fig:anisotropic_CFIFFI_multiE}). The energy spectra of (anti)neutrinos and the collision rates are shown in the left and middle panels of Fig.~\ref{fig:comparison_CCSN_NSM}. In the right panel, we report the values of $\Im(\Omega)_\mathrm{max}$ obtained assuming homogeneity and isotropy, for the multi-energy and the two energy-averaged approaches discussed in Sec.~\ref{subsec:isotropic_CFI}.

The CCSN point is not collisional-unstable, although the monochromatic methods predict a growth rate of a few $10^4 \, \mathrm{s}^{-1}$. The NSM point is unstable, and the growth rate is well estimated by the energy-averaged methods (with a factor $\simeq 2$ between the method B value and the multi-energy one), consistent with our observations in Sec.~\ref{subsec:isotropic_CFI}. The physical difference between these two configurations is clear from the top panels of Fig.~\ref{fig:comparison_CCSN_NSM}: in the NSM case, the density of $\nu_x$ is negligible, while in the CCSN case we have $\N_{xx} \lesssim \overline{\N}_{ee}$. If we can neglect the presence of $\nu_x$, we have shown in the main text that the instability criterion for the monochromatic methods and for the multi-energy analysis are the same [see Eq~\eqref{eq:limit_stability_plus_B} and surrounding text]. Furthermore, the difference in densities between $\nu_e$ and $\bar{\nu}_e$ is smaller in the NSM case, which corresponds to a system further away from stability. Indeed, for gapless modes, there is a maximum asymmetry allowed for the system to be unstable, and the energy-averaged methods perform quantitatively better away from this instability threshold (see Fig.~5 in~\cite{Fiorillo:2025zio} for an illustration).

%%%%%%%%%%%%%%%%%%%%%%%%%%%%%%%%%%%%%%%
\appsection{Other NSM snapshots}
\label{app:other_snapshots}
%%%%%%%%%%%%%%%%%%%%%%%%%%%%%%%%%%%%%%%

In this appendix, we apply our stability analysis to other NSM simulation snapshots, showing that the results obtained in the main text apply more generally.

\subsection{``Early" postmerger phase}

\begin{figure}[!ht]
    \centering
    \includegraphics[width=\columnwidth]{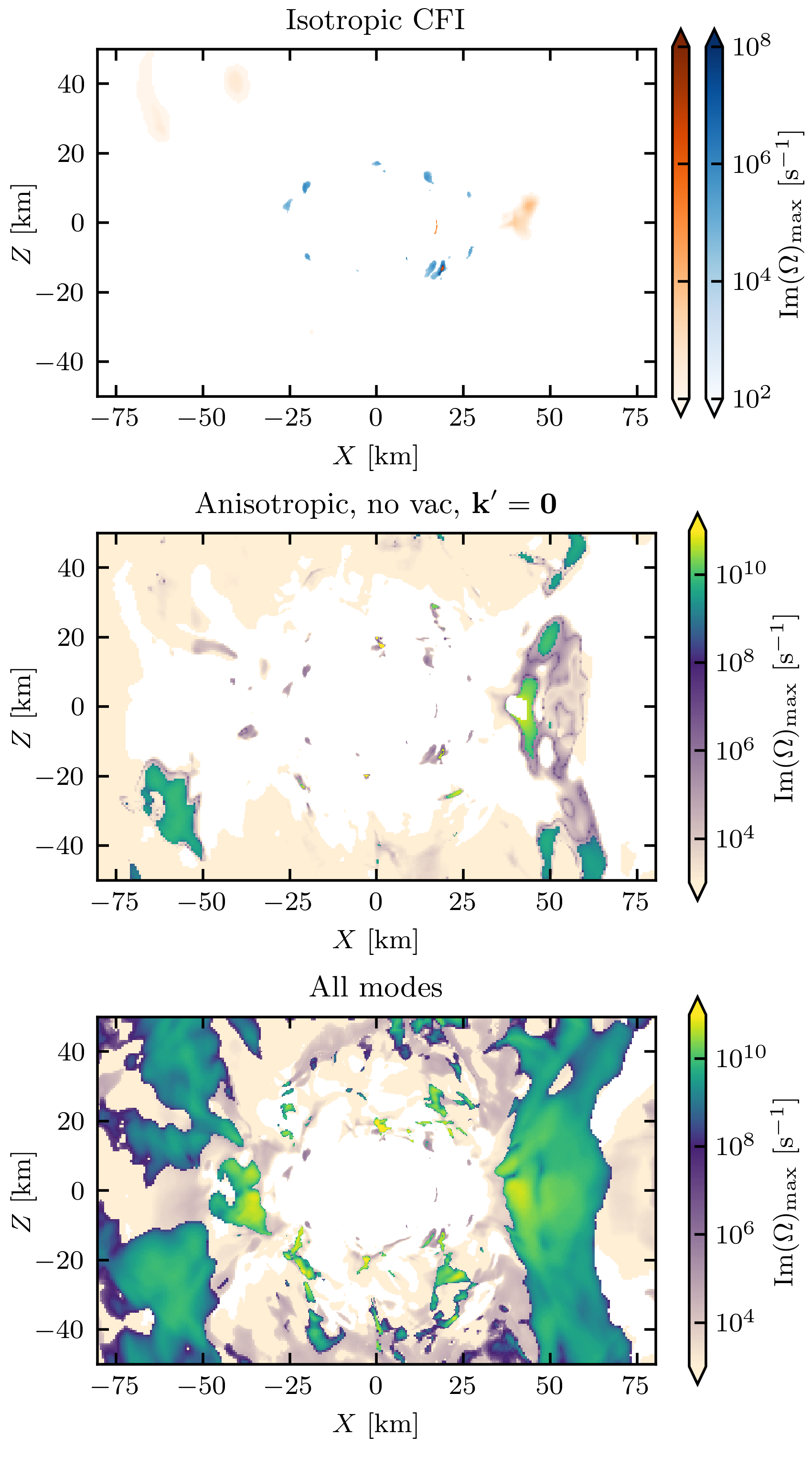}
    \caption{Flavor instabilities in the 3 ms postmerger snapshot from the same NSM simulation studied in the main text. \emph{Top:} results assuming isotropic distributions and without a vacuum term, showing the gapless (orange tones) and gapped (blue tones) CFI regions. \emph{Middle:} results from moment LSA at $\vec{k}'=\vec{0}$, without a vacuum term. \emph{Bottom:} the same as the middle panel but including the vacuum term (NO), and superimposing the FFI estimate~\eqref{eq:approx_FFI}.}
    \label{fig:summary_M13ms}
\end{figure}

We consider here an earlier snapshot from the simulation~\cite{Foucart:2024npn}, namely, the 3 ms postmerger snapshot (while in the main text we focused on the 7 ms one). The results are shown in Fig.~\ref{fig:summary_M13ms}, with from top to bottom: the isotropic CFI (compare with the top left panel of Fig.~\ref{fig:isotropic_CFI_allmethods}), the anisotropic FFI+CFI for the zero mode (compare with Fig.~\ref{fig:anisotropic_momentLSA_multiE}), and the combination of the zero mode moment LSA with vacuum term and the FFI estimate~\eqref{eq:approx_FFI} (compare with Fig.~\ref{fig:allFI_multiE}).

A key difference with the 7 ms postmerger snapshot is that in the 3 ms snapshot there is a system-wide overabundance of $\bar{\nu}_e$ over $\nu_e$, which is consistent with an early phase of protonization. At 7 ms, the system is closer to a slow, quasi-steady-state evolution where some regions are protonizing and others are neutronizing. A consequence of $\overline{\N}_{ee} > \N_{ee}$ is that only gapped modes can occur, and we know they can be described correctly with the monochromatic method A. Nevertheless, we see in Fig.~\ref{fig:summary_M13ms} that the CFI is mostly subdominant, even in the anisotropic case (middle panel), where anisotropy-driven CFIs appear at large $|Z|$, but with very small growth rates. The inclusion of the vacuum term does not dramatically alter the picture, even though they lead to larger growth rates than the CFI at larger distances.

\subsection{Different component masses}

\begin{figure}[!ht]
    \centering
    \includegraphics[width=\columnwidth]{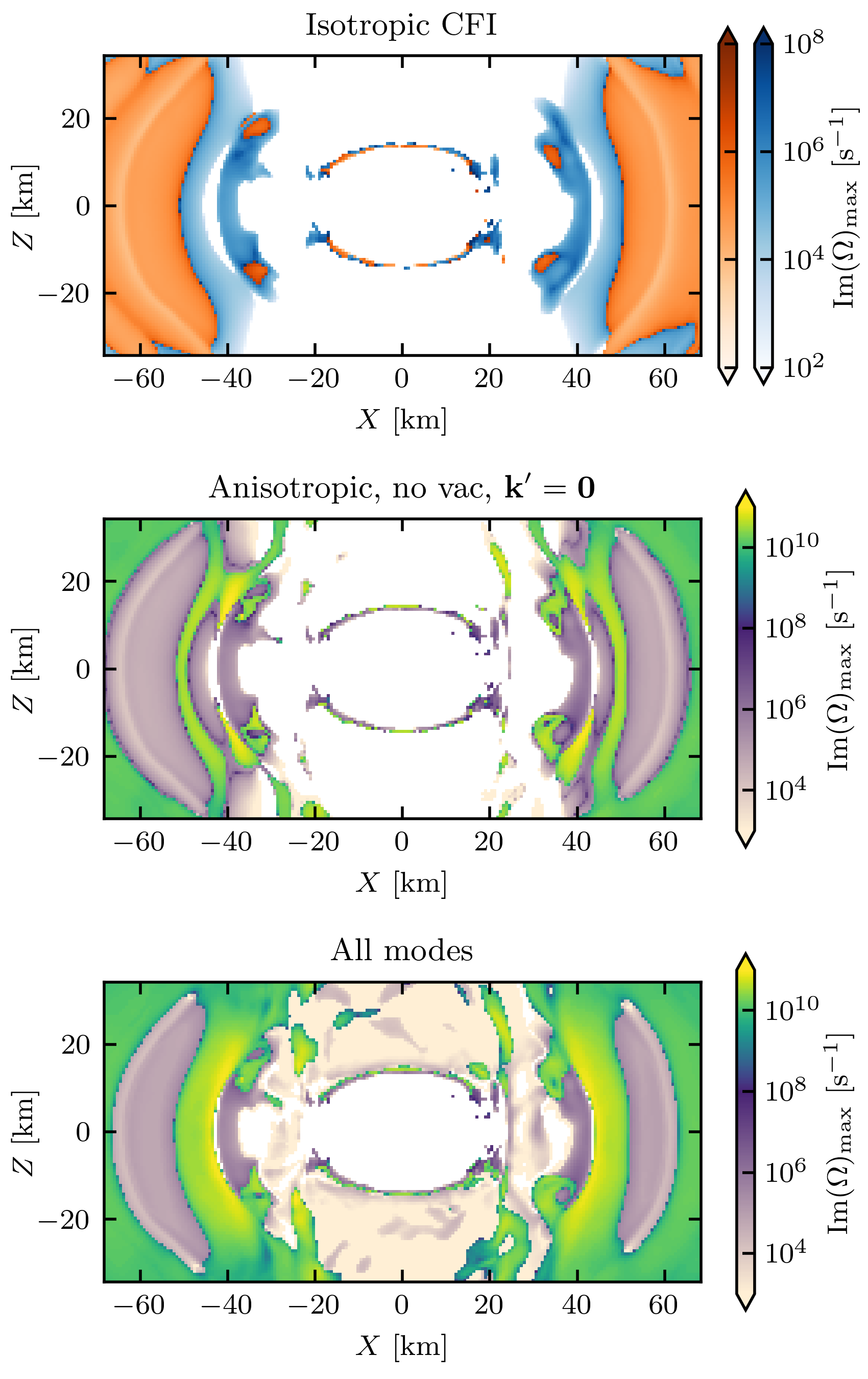}
    \caption{Flavor instabilities in the 5 ms postmerger snapshot from~\cite{Foucart:2016rxm}. Plotting conventions are identical to Fig.~\ref{fig:summary_M13ms}. See the top panel of Fig.~8 in~\cite{Froustey:2023skf} for the FFI-only plot equivalent to the middle panel here, which includes the vacuum term.}
    \label{fig:summary_OldM1}
\end{figure}

Finally, we consider in Fig.~\ref{fig:summary_OldM1} a 5 ms postmerger snapshot from an independent M1 simulation, taken from Ref.~\cite{Foucart:2016rxm}. This simulation of a neutron star merger with component masses of $1.2 M_\odot$ and $1.2 M_\odot$ was performed with
an earlier version of the two-moment radiation transport code SpEC also used in~\cite{Foucart:2024npn}. While in the simulation studied in the main text, the HMNS collapses into a black hole at $t \sim 8.5 \, \mathrm{ms}$, the simulation in~\cite{Foucart:2016rxm} shows no sign of collapse in its $10 \, \mathrm{ms}$ of evolution. However, this snapshot is very similar to the 7 ms one studied in the main text, with thin bands of mostly gapless CFI near the surface of the HMNS, and alternance of gapless and gapped CFI regions in the tidal arms. Although we do not show it here, the qualitative and quantitative agreement of the monochromatic CFI estimate~\eqref{eq:combination_minusA_plusB} is excellent, with once again a ratio of gapless growth rates of about $\sim 2$ between the method B and multi-energy estimates.

Similarly to what we see in Fig.~\ref{fig:allFI_multiE}, when including anisotropies the CFI extends the instability regions in the disk, while slow modes appear in the polar regions. There are also likely slow mode regions at large $|X|$, but the simulation data available do not extend to that region. The instability landscape discussed in this work thus seems to generally extend to ``late" postmerger moment-based simulations, although using snapshots from completely independent simulation codes would be needed to definitely confirm that trend.

%\clearpage

\bibliographystyle{apsrev4-2}
\bibliography{references}

\end{document}